\documentclass[fleqn,usenatbib]{mnras}

\usepackage{newtxtext,newtxmath}

\usepackage[T1]{fontenc}

\DeclareRobustCommand{\VAN}[3]{#2}
\let\VANthebibliography\thebibliography
\def\thebibliography{\DeclareRobustCommand{\VAN}[3]{##3}\VANthebibliography}

\usepackage{graphicx}	
\usepackage{amsmath}	
\usepackage{amssymb}	

\title[Distant {[O III]} Clouds around AGN]{The TELPERION Survey for Distant {[O III]} Clouds around Luminous and Hibernating AGN}

\author[W.C. Keel et al.]{
William C. Keel,$^{1,2}$\thanks{E-mail: wkeel@ua.edu}
Alexei Moiseev,$^{3,4}$
D.V. Kozlova,$^{3}$
A.I. Ikhsanova,$^{3,5}$
D.V. Oparin,$^{3}$
R.I. Uklein,$^{3}$
\newauthor
A.A. Smirnova,$^{3}$
and M.V. Eselevich$^{6}$
\\
$^{1}$Dept. of Physics and Astronomy, University of Alabama, Box 870324, Tuscaloosa, AL 35487, USA\\
$^{2}$SARA Observatory, Embry-Riddle Aeronautical University, Daytona Beach FL 32114, USA \\
$^{3}$Special Astrophysical Observatory, Nizhny Arkhyz, Russia\\
$^{4}$Lomonosov Moscow State University, Sternberg Astronomical Institute, Universitetsky pr. 13, Moscow 119234, Russia\\
$^{5}$Kazan Federal University, Kremlevskaya str.18, 420008 Kazan, Russia\\
$^{6}$Institute of Solar-Terrestrial Physics, Russian Academy of Sciences, Siberian Branch, Irkutsk, Russia
}

\date{Accepted XXX. Received YYY; in original form ZZZ}

\pubyear{2021}

\begin{document}
\label{firstpage}
\pagerange{\pageref{firstpage}--\pageref{lastpage}}
\maketitle

\begin{abstract}
We present a narrowband [O III] imaging survey of 111 AGN hosts and 17 merging-galaxy systems, 
in search of distant extended emission-line
regions (EELRs) around AGN (either extant or faded). Our data reach deeper
than detection from the broadband SDSS data, and 
cover a wider field than some early emission-line surveys used to study extended structure around 
AGN. Spectroscopic followup
confirms two new distant AGN-ionized clouds, in the merging systems 
NGC 235 and NGC 5514, projected at 26 and 75 kpc from the nuclei (respectively). We also
recover the previously-known region in NGC 7252.
These results strengthen the connection between EELRs and tidal features;
kinematically quiescent distant EELRs are virtually always photoionized tidal debris. We see them in $\approx10$\% of 
the galaxies in our sample with tidal tails. Energy budgets 
suggest that the AGN in NGC 5514 has faded by$ >3$ times during the extra light-travel time 
$\approx$250,000 years
from the nucleus to the cloud and then to the observer; strong shock emission in outflows masks 
the optical signature of the AGN. For NGC 235 our data are consistent with but do not 
unequivocally require variation over $\approx$85,000 years. In addition to these very distant ionized clouds, 
luminous and extensive line emission within four
galaxies - IC 1481, ESO 362-G08, NGC 5514, and NGC 7679. IC 1481 shows apparent ionization cones, a rare
combination with its LINER AGN spectrum. In NGC 5514, we measure a 7-kpc shell expanding at $\approx 370$ km s$^{-1}$
west of the nucleus.
\end{abstract}

\begin{keywords}
galaxies: active -- galaxies: interactions -- galaxies: Seyfert
\end{keywords}



\section{Introduction}

Some active galactic nuclei (AGN) are surrounded by regions of ionized gas extended beyond the ordinary narrow-line region
(NLR), extending well into the host galaxies' interstellar medium (ISM) and sometimes even beyond into tidal features or outflows. These
are denoted extended emission-line regions (EELRs; \citealt{Wilson} and \citealt{Canalizo}). Their properties can provide insight into several
facets of AGN - directions of escape of ionizing radiation, the history of the ionizing luminosity, and interactions
with emerging radio jets. Contributions by volunteers with the Galaxy Zoo project \citep{GZ}, using the colour contrast
between ordinary galaxy light and [O III] emission in composite images from the Sloan Digital Sky Survey (SDSS; \citealt{SDSS}),
led to identification of 19 EELRs around low-redshift AGN, most radio-quiet, where about 1/3 showed
energy mismatches between the AGN as observed and the cloud properties suggesting long-term fading over
spans $\approx 10^4$ years \citep{Keel2012}. {Using increasingly distant regions of ionized gas around AGN offers a way to probe their past ionizing luminosity,
and thereby examine their variability over otherwise inaccessible timescales beyond the century or so spanned by our
direct operations, including photographic archives.} Results from this sample motivated us to undertake further surveys in several directions.

To complement continued searches\footnote{Ongoing discussion in https://www.zooniverse.org/projects/zookeeper/galaxy-zoo/talk/1269/858076?comment=2551725} for AGN-ionized clouds by Galaxy
Zoo volunteers using SDSS, DES, and Pan-STARRS1 images, 
we have conducted a narrowband survey in a redshifted [O III] filter
to seek fainter clouds which would be lost against the brighter sky background 
in the survey passbands, over wider regions than were studied in many of the early examinations of EELRs. The study by \citet{Mkn1} became the first
phase of this larger survey, named TELPERION\footnote{Nominally an acronym for Tracing Emission Lines to Probe Extended Regions Ionized by Once-active Nuclei, but more evocative for readers of Tolkien's work on the First Age of Middle Earth, 
where it gives a connotation of vanished brilliance.}. Use of long exposures in narrow filters allows 1m telescopes to reach
deeper in this application than the SDSS broadband images, thereby complementing the \cite{Keel2012} survey which
examined $\approx 15,000$ galaxies to a higher surface-brightness limit. We are particularly interested in very large or 
distant EELR clouds, since these are the ones where light-travel times let us see longest luminosity histories for the AGN.
As in \cite{Keel2012}, we {seek AGN-ionized gas at} projected distances $>10$ kpc from the AGN, roughly marking the distinction between the ISM of the
AGN host galaxy and gas emplaced by additional processes (tidal disturbance or outflows). The circumnuclear ionized regions
around some AGN do extend farther than this, often associated with tidal distortions. The abbreviation EELR has been used to describe
both, so we will call the ones we seek, outside the host ISM, as distant EELRs.

Known distant EELRs, including those associated with fading AGN, occur around nuclei
with a range of currently-observed activity. We see them around bright AGN,
including both ordinary and ``type 2" QSOs, Seyfert galaxies, as well as LINERs and
inactive nuclei \citep{NGC7252}. Accordingly, we sought a balance
between sampling wide ranges of host-galaxy properties and the number of galaxies that could 
feasibly be observed. We observed subsamples selected on merger status, and on
AGN luminosity broadly defined. The survey
recovered some known EELRs, found two new distant cases, and adds to
the statistics of their occurrence around broad categories of galaxies. 
For some features of interest, we have followup spectroscopic data to
examine their ionization source and kinematics, confirming several ionization cones and 
EELR clouds as being AGN-photoionized. Both newly-detected distant EELR clouds are associated with
tidal debris, continuing a pattern seen in previous surveys.

We calculate projected sizes and luminosities using a Hubble constant $\rm H_0 = 70$ km s$^{-1}$ Mpc$^{-1}$,
$\Lambda _M=0.286$, and flat geometry.

\section{Survey Sample, Observations, and Analysis}

\subsection{Samples}

The initial phase of this survey, reported by \citet{Mkn1}, included nearby Seyfert galaxies for which surrounding H I
structures were known. This way, we would know immediately how ionized and neutral structures were spatially
related, and what fractions of extended H I envelopes host distant EELRs. If the outer gas content is independent of the
properties of the AGN, for example, this would tell what fraction of AGN would produce distant EELRs if there were
suitable neutral gas to provide a tracer when ionized.
Building on the success of that phase in finding a new distant EELR around Mkn 1, out of 26 Seyfert galaxies,
we intended this survey to reach surface brightness levels 1/10 that seen in, for example,
Hanny's Voorwerp near IC 2497 \citep{Voorwerp}. For otherwise identical situations, this would allow detection of gas
over three times farther from the AGN and responding to its energy output at epochs three times longer before the
epoch at which we see the AGN directly. We report here observations of two additional galaxy samples - an extended
version of the \citet{Toomre} merger sequence, and a wide-ranging set of AGN host galaxies.

The redshift range for the survey was selected to use off-the-shelf filters for
economy. For redshifted [O III] emission, we used filters denoted F510, from Custom Scientific\footnote{https://customscientific.com/}, 
with designed peak transmission at 510 nm and FWHM 10 nm. They were installed first on the SARA
1m telescope at Kitt Peak, and later at the 1m Jacobus Kapteyn Telescope (JKT) on La Palma,
now operated by the SARA consortium \citep{SARA}.  In the $f/7.5$ beam of 
the Kitt Peak
telescope, we expect a blueward shift of the peak by no more than 8 \AA\ .
The half-transmission points correspond to [O III] $\lambda 5007$ at redshifts 
$=0.0086 - 0.0285$, $cz = 2580-8570$ km s$^{-1}$. We used this redshift range
for our target lists in each survey phase.

Since most known (kinematically-quiescent) EELRs are associated with interacting and merging systems,
we examined all galaxies among the well-known \citet{Toomre} mergers within our redshift and declination limits, augmented by
morphologically similar systems from \citet{KeelWu}. These were observed only from Kitt Peak, as listed in Table \ref{tbl-mergers}.
The table also collects information from the literature on the nuclear spectra, to identify optical AGN signatures.

\begin{table*}
	\centering
	\caption{Log of imaging observation for the extended Toomre merger sequence.
	The UT date of observation is listed for images in redshifted [O III], all from Kitt Peak. As a guide to
	the merger and interaction stage, columns 5-7 indicate how many separate galaxies, nuclei, and tidal tails appear
	in optical images. Column 3 indicates which galaxies were included in the original Toomre sequence. The final column shows the    	
	adopted classification of the nucleus from optical spectroscopy and the source of this information.}
	\label{tbl-mergers}
	\begin{tabular}{lccccccll} 
		\hline
Name & Arp & Toomre? & $z$ & Galaxies & Nuclei & Tails & Date obs (YYYYMMDD) & Nucleus, source\\
\hline
NGC  455   &   164  &            &  0.0176 & 1 & 1 & 2 & 20131202 & Quiescent \citealt{Falco} \\  
NGC 2418   &   165  &           &   0.0169   & 1 & 1 & 2 & 20140106 & Passive, \cite{Bernloehr} \\ 
NGC 2623   &  243 &  Toomre   &0.0182  & 1 & 1 & 2 &  20131202 & LINER \citealt{Keel1984} \\
NGC 2865   &        &                 &   0.0087 & 1 & 1 & 1 &20131115 & E+A \citealt{6dFRS} \\ 
NGC 2914   &   137  &              &   0.0110  & 1 & 1 & 2 &  20140106 & Quiescent SDSS \\
Arp 221    &   221  &                 &   0.0180   &   1 & 2 & 2 & 20140115\\ 
NGC 3303   &   192  &              &   0.0201  & 2 & 2 & 1 & 20170418 &  LINERs \citealt{Keel1985}\\  
NGC 3509   &   335 & Toomre   &   0.0255   & 1 & 1 & 2 & 20140524& LINER \citealt{Falco}\\ 
NGC 3921   &   224 & Toomre   &   0.0200  & 1 & 1 & 2 & 20140515& LINER \citealt{Stauffer}\\ 
NGC 4676   &  242 & Toomre    &   0.0220  & 2 & 2 & 2 & 20140106 & LINERs \citealt{KKHH85} \\ 
IC   883   &   193  &                    &   0.0239  & 1 & 2 & 2 & 20140106 & LINER/composite \citealt{Keel1985}\\  
IC 4553    &   220  &                  &   0.0182 & 1 & 2 & 1 & 20210211 & LINER \citealt{Falco}\\ 
NGC 6240   &        &                  &   0.0242   & 1 & 2 & 3 & 20160430& LINER \citealt{Keel1990}\\ 
NGC 6621/2 &  81 & Toomre    &   0.0208   & 2 & 2 & 1 & 20131002& Starburst+LINER \citealt{Keel1996}\\ 
NGC 7252   & 226 &  Toomre   &     0.0158  & 1 & 1 & 2 & 20150719 & Quiescent \citealt{NGC7252} \\ 
NGC 7585   &   223  &             &   0.0114   & 1 & 1 & 1 & 20131108 &  Quiescent \citealt{Moustakas}\\ 
NGC 7592   & & Toomre &       0.0244   & 2 & 2 & 2 & 20151102 & Composite \citealt{Moustakas}\\ 
		\hline
	\end{tabular}
\end{table*}

We then surveyed a much larger set of AGN host galaxies, regardless of morphology or interaction status. 
Desiring a nearly all-sky sample to best use telescope time, and in view of the limited sky coverage then available for
the SDSS and limited depth for X-ray surveys, we opted for the more miscellaneous but broad selection of AGN in the
VCV catalog \citep{VCV}. In view of our limited knowledge of what properties are associated with the incidence of EELRs, we
used a simple selection on listed VCV absolute magnitude (brighter than $M = -20$, from inhomogeneous sources), plus 
requiring declination north of $-36^\circ$, allowing observations from Kitt Peak for more than one hour per night.
This selection yielded 120 AGN host galaxies, of which 9 were already observed by \citet{Mkn1} and not re-observed.
The remaining galaxies observed in this study are listed in Table \ref{tbl-vcvsample}.
Among these, NGC~ 2329 is listed in VCV with a 1-degree error in declination, and MCG -02-51-008 is misnamed as
MCG~ -5-51-008 (the listed coordinates from VCV correspond to MCG -02-51-008). Both E and W components of NGC 7592 were
included in this sample, and also appear as one combined entry in the Toomre sample. The VCV compilation includes
AGN classifications from a wide range of data in the literature; some of these galaxies may not show spectroscopic
AGN signatures when examined more closely.

Some of the galaxies in our samples have extended emission-line material over very large regions for reasons other than photoionization by an AGN, but their [O III] emission is weak enough to avoid confusion in this survey.
NGC 6240 has a very extensive network of emission-line loops and filaments associated with its starburst wind 
(\citealt{Heckman87}, \citealt{Veilleux2003}, \citealt{Yoshida}, \citealt{AMUSING}) as seen in H$\alpha$ and [N II], but we do not detect [O III] outside the main body of the merger. Likewise the NGC 1275 filament system, variously attributed to shock ionization and
cooling of the intracluster material, has [O III]/H$\beta < 1$ (\citealt{KentSargent}, \citealt{Hatch}). Surveying [O III] rather than, for example, H$\alpha$, immediately reduces the number of non-AGN-photoionized regions detected. 

\begin{table*}
	\centering
	\caption{Object list, and log of imaging observations, for the VCV AGN sample. The absolute magnitude $M$
	is taken from the VCV catalog, as are the AGN types with expansion from their compact notation. The final
	column shows the UT dates each was observed, in the order $V$, F510 when both were not from Kitt Peak (KP) on the
	same night. Data from Cerro Tololo (CT) or the JKT are indicated.}
	\label{tbl-vcvsample}
	\begin{tabular}{rrlclrl} 
		\hline
$\alpha_{2000}$ & $\delta_{2000}$ & Name & $z$ & AGN type & $M$ & Date Observed ($V$, F510) \\
\hline
00 11 06.6 & -12 06 27 & Mkn 938          & 0.019 & Sy 2   & -20.3 & JKT 2016-09-28, KP 2011-11-14 \\
00 18 23.5 & +30 03 47 & NGC 71           & 0.022 & Sy 2   & -20.0 & JKT 2016-09-01, KP 2017-09-19 \\
00 42 52.8 & -23 32 28 & NGC 235          & 0.022 & Sy 1   & -20.6 & JKT 2016-09-28, JKT 2019-11-26 \\
00 55 02.6 & -19 00 17 & ESO 541-G001     & 0.021 & Sy 2   & -21.0 & JKT 2016-09-28, KP 2017-11-14\\
00 59 40.1 & +15 19 51 & UGC 615          & 0.018 & Sy 2   & -20.1 &  CT 2016-11-30, KP 2016-10-09 \\
01 14 07.0 & -32 39 02 & IC 1657          & 0.012 & Sy 2   & -20.4 & CT 2015-12-29, KP 2015-12-29 \\
01 16 03.6 & +04 17 39 & Mkn 565          & 0.021 & Sy?    & -21.2 & CT 2016-11-30, KP 2016-10-12 \\
01 31 50.4 & -33 08 09 & ESO 353-G09      & 0.017 & Sy 2   & -20.6 & JKT 2016-09-28, KP 2017-09-19 \\ 
01 34 26.3 & -15 48 55 & MCG -03.05.007   & 0.020 & Sy     & -20.5 & CT 2016-10-31, KP 2017-11-10 \\ 
01 45 25.4 & -03 49 38 & MCG -01.05.031   & 0.018 & Sy 2   & -20.3 & CT 2016-10-31, KP 2017-09-19  \\
01 52 49.0 & -03 26 48 & MCG -01.05.047   & 0.017 & Sy 2   & -20.1 &  CT 2016-10-31, KP 2016-10-09 \\ 
02 00 14.9 & +31 25 46 & NGC 777          & 0.017 & Sy 2   & -20.6 & KP 2015-10-13\\
02 01 06.5 & -06 48 56 & NGC 788          & 0.013 & Sy 1   & -20.0 & KP 2015-12-10, KP 2017-11-10 \\
02 34 20.1 & +32 30 20 & NGC 973          & 0.016 & Sy 2   & -20.6 & KP 2015-11-02 \\
02 41 38.7 & -28 10 17 & IC 1833          & 0.017 & Sy 2   & -20.0 & KP 2015-11-12 \\
02 42 35.7 & +34 45 46 & NGC 1050         & 0.013 & Sy 2   & -20.2 & KP 2016-03-10 \\
02 49 03.9 & -31 10 21 & IC 1859          & 0.020 & Sy 2   & -20.2 & KP 2015-12-10 \\
03 11 14.7 & -08 55 19 & NGC 1241         & 0.013 & Sy 2   & -21.0 & KP 2015-10-13 \\
03 19 48.2 & +41 30 42 & NGC 1275         & 0.017 & Sy 1.5 & -21.2 & KP 2015-12-29 \\ 
03 42 03.6 & -21 14 37 & ESO 548-G81      & 0.015 & Sy 1   & -20.2 & KP 2015-11-12 \\
04 13 49.7 & -32 00 24 & ESO 420-G13      & 0.012 & Sy 2   & -20.2 & CT 2016-01-08, KP 2016-02-11\\
04 48 37.2 & -06 19 12 & NGC 1667         & 0.015 & Sy 2   & -20.2 & KP 2015-12-10 \\
04 58 54.6 & -00 29 20 & NGC 1713         & 0.015 & LINER  & -20.6 & KP 2016-02-27 \\
05 08 19.7 & +17 21 47 & Zw 468.002       & 0.017 & Sy 2   & -20.8 & JKT 2016-10-29, KP 2016-11-25 \\
05 10 13.9 & -29 24 13 & MCG -05.13.010   & 0.023 & LINER  & -20.0 & CT 2015-12-29, KP 2015-12-29 \\
05 11 09.0 & -34 23 36 & ESO 362-G08      & 0.016 & Sy 2   & -20.8 & CT 2015-12-29, KP 2016-02-17 \\
06 23 46.5 & -32 13 00 & ESO 426-G002     & 0.022 & Sy 2   & -20.5 & CT 2016-01-08, KP 2016-02-11\\
06 23 51.0 & -23 11 37 & MCG -04.16.001   & 0.024 & Sy     & -20.7 &  CT 2016-04-08, KP 2016-11-25 \\
06 32 47.5 & +63 40 23 & UGC  3478        & 0.012 & NLSy 1 & -20.6 & KP 2016-11-12 \\
06 47 45.8 & +74 28 54 & NGC 2258         & 0.014 & LINER  & -20.9 & KP 2016-02-17\\
06 54 35.1 & +50 21 10 & Mkn 373          & 0.020 & Sy 1   & -20.4 & KP 2016-02-27 \\
07 09 08.0 & +48 36 55 & NGC 2329         & 0.019 & LINER  & -21.3 & KP 2016-03-10 \\ 
07 42 32.9 & +49 48 35 & Mkn 79           & 0.022 & Sy 1.2 & -20.0 &  KP 2015-11-12 \\
07 44 09.1 & +29 14 49 & UGC 3995         & 0.015 & Sy 2   & -20.6 & JKT 2016-10-29, KP 2017-03-02 \\
07 59 40.2 & +15 23 15 & UGC 4145         & 0.016 & Sy 2   & -20.1 & KP 2016-04-01 \\
08 07 41.0 & +39 00 13 & Mkn 622          & 0.023 & Sy 2   & -20.1 & JKT 2017-01-17, KP 2016-02-11 \\
08 08 47.1 & +00 18 01 & RX J08087+0018   & 0.019 &        & -20.4 & CT 2016-04-08, KP 2017-03-02 \\
08 35 38.8 & -04 05 19 & NGC 2617         & 0.014 & Sy 1.8 & -20.0 & KP 2016-03-09 \\
08 37 26.6 & +40 02 07 & UGC 4498         & 0.024 & Sy?    & -20.3 & KP 2016-03-01 \\
08 37 32.7 & +28 42 18 & NGC 2619 & 0.012 & Sy 1   & -20.3 & JKT 2017-01-18, KP 2017-03-02 \\ 
08 43 38.0 & +50 12 21 & NGC 2639         & 0.011 & LINER  & -20.3 & JKT 2017-01-17, KP 2017-03-30 \\
09 19 02.3 & +26 16 11 & NGC 2824         & 0.008 & Sy?    & -21.8 & JKT 2017-01-17, KP 2017-03-30 \\ 
09 20 46.2 & -08 03 21 & MCG -01.24.012   & 0.020 & Sy 2   & -20.8 & CT 2016-04-08, KP 2016-05-03 \\ 
09 34 43.8 & -21 55 42 & ESO 565-G19      & 0.015 & Sy 2   & -20.2 & CT 2016-02-06, KP 2016-02-11 \\
09 37 41.1 & -22 02 06 & NGC 2945         & 0.016 & Sy     & -21.0 & CT 2016-11-11, KP 2017-03-02 \\
09 50 56.5 & -04 59 06 & MCG -01.25.049   & 0.022 & Sy 2   & -20.3 & CT 2016-11-30, KP 2017-04-05\\
09 51 55.0 & -06 49 23 & NGC 3035         & 0.015 & Sy 1.8 & -20.4 & JKT 2017-01-18, KP 2017-04-18 \\
10 00 40.9 & -31 39 51 & PKS 0958-314     & 0.009 & LINER  & -20.7 & KP 2016-03-10 \\ 
10 02 07.0 & +03 03 27 & IC 588 & 0.023 & Sy 1   & -20.1 & JKT 2017-01-19 KP 2017-04-05 \\ 
10 37 00.1 & +18 08 08 & NGC 3303        & 0.020 & LINER  & -20.0 & JKT 2017-01-19, KP 2017-04-20 \\
10 39 46.3 & -05 28 59 & MCG -01.27.031   & 0.021 & Sy 1   & -20.7 & CT 2016-04-08, KP 2017-04-05 \\
11 07 18.1 & -19 28 17 & NGC 3497         & 0.012 & LINER  & -20.0 & CT 2016-02-06, KP 2016-02-11 \\
11 10 48.0 & -28 30 04 & ESO 438-G09      & 0.024 & Sy 1   & -20.4 & CT 2016-04-08, KP 2016-03-02 \\
11 24 02.8 & -28 23 15 & IRAS 11215-2806  & 0.014 & Sy 2   & -20.9 & CT 2016-04-08, KP 2017-03-30\\
11 27 23.4 & -29 15 27 & ESO 439-G09      & 0.023 & Sy 2   & -20.2 & CT 2016-04-08, KP 2016-04-01 \\
11 35 57.5 & +70 32 07 & NGC 3735         & 0.009 & Sy 2   & -20.2 & KP 2016-03-09 \\
11 40 13.9 & +24 41 49 & NGC 3798         & 0.012 & Sy 1   & -20.4 & JKT 2016-05-05, KP 2016-05-03 \\
11 42 11.3 & +10 16 40 & NGC 3822         & 0.020 & Sy 1.9 & -20.8 & JKT 2016-05-05, KP 2016-05-05 \\
11 43 18.6 & -12 52 41 & NGC 3831         & 0.018 & Sy 2   & -20.7 & CT 2017-02-09, KP 2017-02-27 \\
11 45 11.6 & -09 18 51 & NGC 3858         & 0.019 & Sy 2   & -20.5 & CT 2017-02-09, KP 2017-04-18 \\
11 46 12.2 & +20 23 29 & NGC 3884         & 0.023 & LINER  & -21.2 & KP 2016-04-01 \\
11 55 40.1 & -12 01 39 & NGC 3974         & 0.019 & LINER  & -20.0 & CT 2017-02-09, KP 2017-03-02\\
11 55 57.4 & +06 44 56 & NGC 3976         & 0.008 & Sy 2   & -20.4 & KP 2016-02-07 \\
\hline
\end{tabular}

\end{table*}

\begin{table*}
\contcaption{VCV AGN sample}
\label{tbl-vcvsample:continued}
	\begin{tabular}{rrlclrl} 
\hline
$\alpha_{2000}$ & $\delta_{2000}$ & Name & $z$ & AGN type & $M$ & Observed\\
\hline
12 04 43.4 & +31 10 38 & UGC 7064         & 0.024 & Sy 1.9 & -20.2 & JKT 2017-01-19, KP 2017-04-18 \\
12 12 18.9 & +29 10 46 & NGC 4169         & 0.013 & Sy 2   & -20.6 & JKT 2016-05-6, KP 2016-05-05 \\
12 21 49.0 & -24 10 05 & ESO 506-G04      & 0.013 & LINER  & -21.1 & CT 2017-01-28, KP 2017-03-30\\
12 26 16.1 & -07 40 50 & NGC 4404         & 0.018 & LINER  & -21.2 & JKT 2017-02-04, KP 2017-04-20 \\
12 38 54.6 & -27 18 28 & ESO 506-G27      & 0.024 & Sy 2   & -20.2 & CT 2016-04-26, KP 2017-04-05 \\
12 41 44.7 & +35 03 45 & NGC 4619         & 0.023 & Sy 1.0 & -21.5 & KP 2016-03-10 \\
12 51 31.8 & -26 27 07 & MCG -04.30.029   & 0.011 & LINER  & -20.1 & CT 2017-01-28, KP 2017-02-27 \\
12 52 36.4 & -21 54 38 & ESO 575-IG016    & 0.023 & Sy 2   & -20.1 & CT 2017-02-29, KP 2017-04-06 \\
13 01 22.8 & -30 56 06 & NGC 4903         & 0.016 & Sy 2   & -21.7 & CT 2017-01-28, KP 2017-03-30 \\
13 06 17.3 & +29 03 47 & NGC 4966         & 0.024 & LINER  & -20.8 & JKT 2017-02-4, KP 2017-04-18 	\\ 
13 08 41.8 & -24 22 57 & PKS 1306-241     & 0.014 & Sy 2   & -20.0 & CT 2017-01-28, KP 2017-04-06 \\
13 10 17.3 & -07 27 15 & MCG -01.34.008   & 0.022 & Sy 2   & -20.3 & JKT 2017-02-04, KP 2017-04-18 \\
13 19 50.0 & -27 24 36 & NGC 5078         & 0.007 & LINER  & -20.6 & CT 2017-01-31, KP 2017-04-05 \\
13 22 24.5 & -16 43 42 & MCG -03.34.064   & 0.017 & Sy 1   & -20.4 & JKT 2017-03-06, KP 2017-04-06 \\
13 24 50.3 & -30 18 28 & NGC 5124         & 0.014 & LINER  & -20.8 & CT 2017-01-31, KP 2017-03-30\\
13 33 26.0 & -34 00 56 & ESO 383-G18      & 0.013 & Sy 1.8 & -20.1 & CT 2017-01-31, KP 2017-04-06 \\
13 36 08.5 & -08 29 52 & NGC 5232         & 0.023 & Sy?    & -21.4 & JKT 2017-02-04, KP 2017-03-02 \\
13 39 22.5 & -32 13 27 & MCG -05.32.059   & 0.024 & LINER  & -21.0 & CT 2016-08-04, KP 2017-03-02  \\
13 40 19.7 & -23 51 28 & NGC 5260         & 0.022 & Sy 2   & -21.3 & CT 2016-04-26, KP 2017-02-27 \\
13 47 40.1 & -30 56 21 & NGC 5292         & 0.015 & LINER  & -21.1 & CT 2016-04-26, KP 2016-05-03 \\
13 50 36.0 & +33 42 17 & NGC 5318         & 0.014 & Sy?    & -20.4 &JKT 2016-05-05, KP 2016-06-17  \\
13 53 26.7 & +40 16 59 & NGC 5353         & 0.008 & Sy?    & -21.2 & JKT 201608-31, KP 2016-06-17  \\
13 58 38.0 & +37 25 28 & NGC 5395         & 0.012 & Sy 2   & -20.3 & KP 2016-04-01 \\
14 01 58.4 & -25 32 25 & ESO 510-G46      & 0.021 & Sy 2   & -20.4 & CT 2016-04-26, KP 2017-04-05 \\
14 13 38.9 & +07 39 33 & NGC 5514         & 0.024 & LINER  & -21.4 & JKT 2016-05-06, JKT 2021-03-18 \\
14 28 31.9 & +27 24 31 & NGC 5635         & 0.014 & LINER  & -20.0 & JKT 2016-05-03, KP 2016-05-03 \\ 
14 47 33.5 & -20 28 02 & MCG -03.38.010   & 0.023 & LINER  & -20.0 & CT 2016-08-05. KP 2016-07-02 \\
15 06 50.7 & -14 34 18 & NGC 5849         & 0.024 & LINER  & -20.1 & CT 2017-03-30, KP 2017-03-30 \\
15 15 23.2 & +55 31 05 & NGC 5905         & 0.011 & H II   & -20.6 & KP 2016-02-27 \\
15 46 16.5 & +02 24 56 & NGC 5990         & 0.013 & Sy 2   & -20.4 & KP 2016-05-05 \\
16 32 32.1 & +82 32 17 & NGC 6251         & 0.024 & Sy 2   & -21.1 & JKT 2016-05-06, KP 2016-06-17\\
16 41 27.8 & +57 47 01 & NGC 6211         & 0.020 & Sy 2   & -20.2 & KP 2016-02-27 \\
16 52 58.9 & +02 24 01 & NGC 6240         & 0.024 & LINER  & -20.7 & KP 2016-04-30 \\
17 20 29.5 & +50 22 37 & NPM1G +50.0375   & 0.024 & Sy 2   & -20.8 & JKT 2016-06-04, KP 2016-06-17\\
20 17 06.3 & -12 05 51 & MCG -02.51.008   & 0.020 & LINER  & -21.2 & CT 2016-08-03, KP 2019-10-01 \\ 
20 27 46.0 & -03 04 37 & NGC 6915         & 0.019 & LINER  & -22.0 & CT 2016-08-05, KP 2016-07-02 \\
20 34 31.4 & -30 37 29 & PGC 64989        & 0.019 & Sy 1   & -21.1 & CT 2016-08-05, KP 2016-10-07 \\
20 35 56.3 & -25 16 48 & NGC 6936         & 0.019 & LINER  & -20.6 & CT 2016-08-05, KP 2016-11-09 \\
20 46 37.2 & -23 37 49 & MCG -04.49.001   & 0.020 & LINER  & -20.0 & CT 2016-08-05, KP 2016-10-09 \\
20 47 34.1 & +00 24 41 & UGC 11630        & 0.012 & Sy 2   & -20.7 & KP 2015-10-13,  KP 2016-07-02 \\
22 00 21.6 & -13 08 50 & IC 1417          & 0.018 & Sy 2   & -20.3 & KP 2015-11-12 \\
22 09 07.7 & -27 48 34 & NGC 7214         & 0.023 & Sy 1.2 & -20.3 & KP 2015-11-02, KP 2016-07-02 \\
22 12 31.5 & +38 40 55 & UGC 11950        & 0.020 & LINER  & -20.2 & JKT 2016-08-31, KP 2016-09-05\\
22 27 05.9 & +36 21 41 & UGC 12040        & 0.021 & Sy 1.9 & -20.1 & JKT 2016-098-31, KP 2016-09-05\\
22 31 20.7 & +39 21 30 &  UGC 12064   & 0.017 & Sy     & -20.2 & JKT 2016-08-31, KP 2016-10-07 \\ 
22 36 03.5 & +33 58 33 & NGC 7319         & 0.022 & Sy 2   & -20.2 & KP 2015-11-12 \\
23 03 15.6 & +08 52 26 & NGC 7469         & 0.017 & Sy 1.5 & -20.9 & JKT 2016-08-31, KP 2017-09-19 \\
23 17 12.1 & -06 54 42 & NGC 7596         & 0.024 & Sy     & -20.0 & JKT 2016-09-01, KP 2016-10-07 \\
23 18 16.3 & +06 35 09 & NGC 7591         & 0.017 & Sy     & -20.8 & JKT 2016-09-01, KP 2016-11-09 \\
23 18 21.8 & -04 24 56 & NGC 7592E        & 0.024 & H II   & -20.1 & V JKT 8/31/16\\
23 18 22.1 & -04 24 59 & NGC 7592W        & 0.024 & Sy 2   & -20.9 & KP 2015-11-02 \\
23 19 25.3 & +05 54 21 & IC 1481          & 0.020 & LINER  & -20.1 & JKT 2016-09-1, KP 2016-11-15 \\
23 29 03.9 & +03 32 00 & NGC 7682         & 0.017 & Sy 1   & -20.0 & JKT 2016-09-01, KP 2016-10-07 \\
23 30 47.7 & -13 29 08 & IC 1495          & 0.021 & Sy 2   & -21.7 & JKT 2016-09-27, KP 2016-11-25 \\
23 41 47.3 & -03 40 02 & MCG -01.60.021   & 0.023 & LINER  & -20.5 & JKT 2016-09-27, KP 2016-11-09 \\
23 56 03.9 & -00 59 18 & IC 1515          & 0.023 & Sy 2   & -20.6 & JKT 2016-09-28, KP 2016-11-09 \\
		\hline
	\end{tabular}
\end{table*}

\subsection{Imaging observations}
Our imaging observations mostly spanned the period 2015 October - 2017 November, with a few later repeats for confirmation
or improved image quality. We used primarily the SARA 1-m telescope at Kitt Peak, Arizona for the
narrowband images, aiming for a single one-hour exposure wherever possible.
Since the image quality is generally better at the JKT than at the SARA telescope on
Kitt Peak, we re-observed confirmed
EELRs with that setup for a better view of their structures once a matching filter and autoguiding were available there.
The field sizes were 14.5 (Kitt Peak) or 11.6 (JKT) arcmin square, with pixel scales 0.44 and 0.34\arcsec per pixel respectively. 
For a centered target, the inscribed circle of complete coverage in Kitt Peak images
projects to radii 77--250 kpc at our sample's minimum and maximum redshifts. As in the Galaxy Zoo survey
\citep{Keel2012}, our primary interest is in EELRs projected more than 10 kpc from the AGN (distant EELRs), both to provide
long differential light-travel times to probe the history of the AGN luminosity, and to examine gas outside
the normal interstellar medium (ISM) of the host galaxy.




Exposures in the F510 filter for redshifted [O III] emission were 
typically 1 hour, at Kitt Peak or the JKT.
The narrowband Kitt Peak images were often affected by internal stray light 
(probably due to an LED associated with the fiber spectrograph, since fixed) which
gave a diagonally sloping background  across the field of view. While the
problem existed, it was sometimes dealt with using similar exposures obtained with
the telescope in the same position on a cloudy night, with the CCD shutter open
but mirror cover closed. A few objects in the Toomre sample were sufficiently affected to produce
flat-fielding artifacts. In any case, median filtering with large kernel sizes generally
suppressed the background structure enough to seek features of the sizes we sought.
A few images were repeated once the problem was repaired. 

\begin{figure*}
\includegraphics[width=155.mm,angle=0]{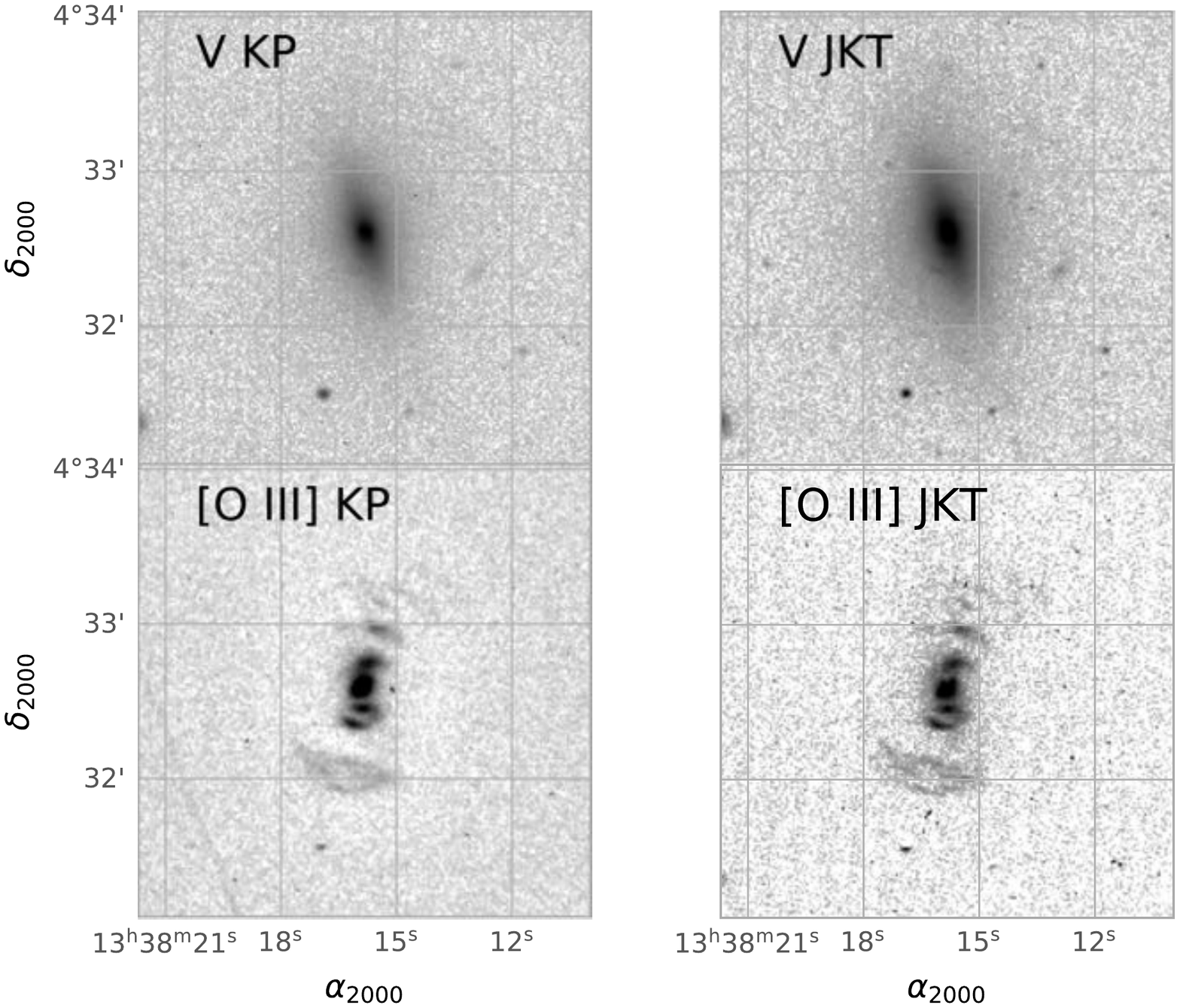}
\caption{Kitt Peak and JKT comparison images of the well-known EELR features in NGC 5252, illustrating the depth of our survey imaging. From Kitt Peak, the
F510 image is the mean of 2 30-minute exposures, while the $V$ image is the mean of 
2 10-minute exposures; from the JKT, we have a single one-hour narrowband exposure and 2 10-minute $V$ exposures.
A $3 \times 3$-pixel median filter was used to smooth and reduce charged-particle artifacts in the continuum-subtracted
[O III] images. Images are shown in an offset-logarithmic scale closely approximating the usual
SDSS image displays.
Effects of filter mismatch are seen in the slight oversubtraction of the smooth galaxy light
in its inner regions, while the [O III] structure is a good match for that seen with
narrower filters (\citealt{Tadhunter}, \citealt{Tsvetanov}) Emission features are detected to a radius 51\arcsec (27 kpc).
NGC 5252 itself was not included in the TELPERION sample because its listed absolute
magnitude in VCV is slightly too faint (-19.6, while our sample limit was -20.0). North is at the
top and east to the left.}
\label{fig-ngc5252}
\end{figure*}

Continuum subtraction used the Johnson $V$ band, sometimes at a different SARA telescope
than the narrowband image to use time more efficiently; for a few galaxies in the Toomre
sample, we used $V$ images obtained with the 1.1-m Hall telescope at Lowell Observatory by 
\cite{KeelWu}.
The goal was for the 
broadband image to have at least twice the signal-to-noise ratio (SNR) on the galaxy as in
the narrowband image, to minimize loss of SNR in continuum subtraction. The quality of
this subtraction was generally limited by PSF matching; seeing, guiding behavior,
and focus shifts sometimes made the subtraction unreliable very close to
galaxy nuclei as well as star images. Since our main goal was detecting large-scale
emission features (targeting clouds more than 10 kpc in projection from the nucleus), this was more of 
an aesthetic than scientific problem. Using the  conservative approach of measuring the compromised regions around stars
brighter then the galaxy nuclei, the largest such region has a radius of 10\arcsec, with a sample median of 6\arcsec. Fig. \ref{fig-ngc5252}
illustrates the depth of the data, comparing broad-band $V$ and
continuum-subtracted [O III] images of the well-known EELRs in NGC 5252 using data matching our survey
exposures.

World-coordinate system (WCS) metadata for each image were determined by the
{\tt astrometry.net} web service (Lang et al. 2010), separately for continuum and narrowband images when they were
from different telescopes or observing sessions.
The $V$ images were resampled to
match the nominal pixel grids of the narrowband images. A simple Gaussian matching 
was initially used to account for PSF differences, with small pixel-level shifts 
sometimes needed when one PSF had an asymmetric core. Relative scaling of the broadband
image started using count rates of foreground stars, and was iteratively changed for
better cancellation of galaxy light (sometimes requiring judgment as to what parts should
be best cancelled when colour gradients are significant). We applied various kinds of smoothing - median 
over $3 \times 3$ or $5 \times 5$ pixels, Gaussian blurring - to seek fainter emission in the presence of residual cosmic-ray effects.
Emission from H II regions in spiral arms, especially in the outer parts of some hosts, was frequently detected, 
and is characterized by alignment along the arms and concentration to bright spots, in contrast to the more diffuse or filamentary appearance of EELR emission.

Somewhat unexpectedly, having long times elapse between narrowband and continuum images led to multiple
asteroid trails which could mimic emission-line features in difference images, similar to the apparent jet of NGC 3303 in the \citet{Arp}
atlas which could not be reobserved and was eventually identified as an unfortunately placed and timed asteroid \citep{Kanipe}. 
For the record, we recorded asteroid (59585) in front of NGC 3822 on 2016-05-5, (43336) near NGC 7591 on 2016-11-09,
and (20865) near the EELR of NGC 5514 during a Kitt Peak reobservation on 2019-05-19.
We caution other galaxy observers.

Comparison with detections and nondetections in followup spectroscopy indicates that our detection limit for 
diffuse [O III] $\lambda 5007$
emission regions is at a typical surface brightness $1.0 \times 10^{-16}$ ergs cm$^{-2}$ s$^{-1}$ arcsec$^{-2}$.
This is about four times deeper than the SDSS-based selection used in the Galaxy Zoo survey reported by 
\cite{Keel2012}, where the peak surface brightnesses even when averaged over the tabulated $2 \times 6.2$\arcsec regions are all
greater than $3.5 \times 10^{-16}$ ergs cm$^{-2}$ s$^{-1}$ arcsec$^{-2}$.

Given the small incidence of distant EELRs, false positives will be much more frequent than
false negatives near the detection threshold. After initial experience with spectroscopic followup.
we first repeated the narrowband imaging observations of candidate diffuse [O III] objects, 
usually with longer exposures, and carried objects through for spectroscopy only when confirmed in this way.
This proved to be effective - all of the [O III] objects confirmed by repeat imaging were
also confirmed as emission-line regions (although not necessarily AGN-ionized EELRs) when
spectroscopically observed. This proved to be a powerful strategy when the observations 
were spread over many observing sessions, as ours were, allowing this kind of feedback and 
refinement.

\subsection{Tunable-filter imaging}

Several of the most interesting systems from this survey were observed in [O III] and the adjacent continuum using
tunable-filter systems. Data for IC 1481 and NGC 5514 used the
tunable-filter imaging mode of the MaNGaL system \citep{MANGAL} at the 2.5m telescope\footnote{http://lnfm1.sai.msu.ru/kgo/main.php} of the Caucasus Mountain Observatory (\citealt{KGO}, \citealt{KGOoptics})
operated by the Shternberg Astronomical Institute. These data 
improved both the typical image quality and sensitivity over our SARA 1m images. These observations were done with the scanning Fabry-Perot interferometer for a FWHM resolution near 13 \AA.  For NGC 5514 (observed on 2018-04-11), on-band and off-band exposures were each 1600 seconds; for IC 1481, observed on 2017-11-17,
the on-band exposure was 4800 seconds accompanied by 2400 seconds in the continuum band. 

Similarly, NGC 7679 was observed at the Russian 6-m telescope in the imaging mode of the SCORPIO-2 focal reducer \citep{SCORPIO2}on 2017-11-25. We used two narrowband (FWHM$\approx35$ \AA) filters centered on redshifted [OIII] $\lambda$5007 
and adjacent continuum, with total exposures 1200 s in the each filter. The pixel scale was 0.35'\arcsec and field of view was 6\arcmin .

\subsection{Archival Hubble Space Telescope data}

A {\it Hubble Space Telescope} image exists for NGC 235, a single 500s exposure from snapshot program 5479 \citep{MalkanGorjianTam}. After interactive patching of the numerous cosmic-ray events in this region of the image and
smoothing, part of the EELR is
(just) detected. The response curve of the F606W filter on WFPC2 indicates that this detection is about equally sensitive to
the wavelengths of [O III] and H$\alpha$+[N II].

\subsection{Long-slit spectroscopy} 

To explore the nature of extended emission-line structures, and identify which ones are photoionized by AGN, we
obtained followup optical spectroscopy. Most of these observations were obtained with the SCORPIO \citep{SCORPIO} and
SCORPIO-2  \citep{SCORPIO2} multimode focal-reducer
systems at the 6-meter telescope (BTA) of the Special Astrophysical Observatory, Russian Academy of Sciences, supplemented by several
observations using the new ADAM multimode instrument at the 1.6-meter telescope of Sayan Observatory, Siberian Branch of the Russian Academy of Sciences \citep{ADAM}. Each instrument uses reimaging optics and a volume-phase holographic grating \citep{VPH} for 
high efficiency and an appropriate match between pixel and image scales. The spectral range and spectral resolution were 3650--7750 \AA\  and $\approx 10$ \AA\  for the SCORPIO observations of IC 1481, 3800-- 8400 and $\approx 4$ \AA\ for NGC 7679, and about 3600-7200 \AA\  and 5 \AA\ for the other SCORPIO-2 data. The ADAM spectra were obtained in the range 3650-7300 \AA\ with a typical resolution 10--15 \AA\  depending on the slit width.

The spectroscopic observations are listed in Table \ref{tbl-spectra}. In most cases, the slit was oriented to pass through the
galaxy nucleus and distant [O III] feature. The companion south of NGC 2329 was observed; weather did not allow proper
intensity calibration for this spectrum as well as NGC 3798, although the important line ratios for classification are scarcely affected. 
For all other data, spectrophotometric calibration used standard stars observed each night.
For NGC 235, at declination $-23^\circ$ and necessarily observed at large air mass, 2-pixel on-chip binning along the slit was used to increase the SNR of the spectrum.

\begin{table}
	\centering
	\caption{Summary of long-slit spectra. Slit widths are in arcseconds, position angles are in degrees north through east,
	and exposure durations are in seconds. 
	All slit locations went through the galaxy nuclei except for NGC 235, acquired best using the SW companion NGC 232, and
	the NGC 5514
	cloud spectrum, which was offset to follow the long axis of the distant emission-line structure.}
	\label{tbl-spectra}
	\begin{tabular}{lrcrcl} 
		\hline
Object & PA$^\circ$ & Date   & Exp & Slit  & Instrument \\
\hline
IC 1481               & 315 & 2017-08-16  & 2400 & 1.0\arcsec & SCORPIO \\ 
MCG -01-25-049 &130  & 2018-03-13 & 1800 &  1.0\arcsec & SCORPIO-2 \\
Mkn 373               & 17  & 2017-12-09 & 2700 & 1.0\arcsec & SCORPIO-2 \\
NGC 235             & 54 & 2020-11-10 & 2400  & 1.5\arcsec & SCORPIO-2 \\
NGC 1050         & 195 & 2018-01-25 & 2400 & 1.0\arcsec & SCORPIO-2 \\ 
NGC 2329 S    & 0 & 2017-12-14 & 2400 & 1.0\arcsec & SCORPIO-2 \\ 
NGC 2617       & 155 & 2018-03-12  & 2700 & 1.0\arcsec & SCORPIO-2 \\ 
NGC 5514 nuc    & 88  & 2018-03-13 & 3600 & 1.0\arcsec & SCORPIO-2 \\
NGC 5514 cloud  & 10.5 & 2020-03-02 & 6000 & 1.0\arcsec & SCORPIO-2 \\
NGC 5635       & 55       & 2018-03-13 & 4500 & 1.0\arcsec & SCORPIO-2 \\
NGC 7679        & 90  & 2017-11-25 & 6000 & 0.5\arcsec & SCORPIO-2 \\ 
Mkn 373        & 17  & 2017-03-19 & 3600 & 3.0\arcsec & ADAM \\
NGC 2617-1    & 334    &  2017-03-21  & 1200 &  2.0\arcsec &  ADAM \\ 
NGC 2617-2      & 333   &   2017-03-27 & 1200 & 2.0\arcsec &  ADAM \\ 
NGC 3798     & 264  & 2017-03-18 & 5400 & 3.0\arcsec &  ADAM \\  
NGC 5514       &  88 & 2017-03-21 & 2400 & 3.0\arcsec &  ADAM \\
NGC 6251       & 170 & 2017-03-16 & 6000 & 2.0\arcsec &  ADAM \\
		\hline
	\end{tabular}
\end{table}

We augment our long-slit spectrum of the cloud in NGC 235 with the optical spectrum of its AGN
obtained by the BAT AGN Spectroscopic Survey (BASS, distinct from the BASS portion of the Legacy Survey) collaboration \citep{BASSDR1}, and adopt their emission-line fit parameters.

Our long-slit spectra allowed a closer examination of some ambiguous AGN types from the VCV catalog.
NGC 2617 is a Sy 1, NGC 5635 a LINER, and NGC 1050 shows nuclear star formation. A candidate emission cloud near NGC 6251 was not detected
spectroscopically, indirectly showing the surface-brightness limits of our [O III] search. Most of the candidate EELR clouds
are shown to be star-forming regions; the BPT classification allows the
possibility that the emission 11\arcsec (3.3 kpc)  NW of the nucleus in NGC 2617, coincident with a segment of spiral arm, is AGN-ionized.

\section{Results}

\subsection{Extended and distant emission-line clouds}

We were able to obtain spectra of all plausible candidate distant EELRs, some of which proved to be
other kinds of emission-line features. Our primary classification tools for ionization mechanisms were the traditional BPT diagrams 
\citep{BPT}, incorporating revised region boundaries from \cite{Kewley2001} and \cite{Kauffmann}. This approach was further 
augmented by the He II/H$\beta$ ratio as described by \cite{Mkn1}, to allow identification of
AGN-ionized gas even at low metallicity, where the strong-line ratios approach those of the starburst
branch of the primary BPT diagram \citep{Groves2006}. 

We identified two new distant AGN-ionized EELR clouds,
near NGC 235 and NGC 5514, and recovered the previously-known cloud in NGC 7252 \citep{NGC7252}. The feature in NGC 235 is
marginally shown (in a single contour) by \cite{Mulchaey}, but not discussed; they analyze the [O III] blobs 10\arcsec 
SW of the nucleus. The EELR in NGC 7252 is centered about 5 kpc from the nucleus in projection and barely extends outward to our 
minimum search radius of 10 kpc, so we do not always include it in the sample comparisons which follow. 

\begin{figure}
\includegraphics[width=90.mm,angle=0]{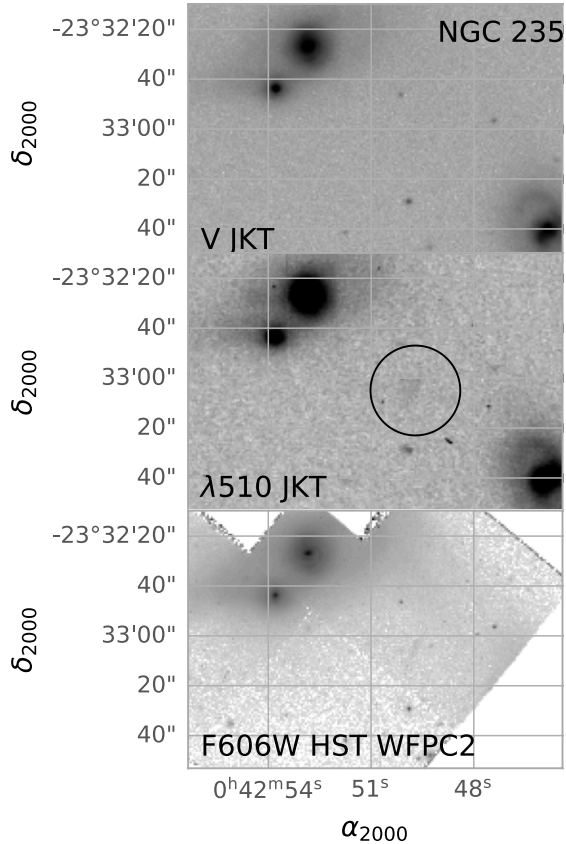}
\caption{The emission-line cloud in NGC 235 (larger galaxy at top left). The figure compares a $V$-band image to show 
the stellar
structure of this interacting system, a view of the cloud as seen in the  F510 narrowband image, and the archival HST WFPC2 F606W image which has been partially corrected for cosmic-ray events using the interactive {\it imedit} tool within IRAF. Each image is shown with
on an offset-log intensity scale, similar to sinh mapping. The HST image has been smoothed with a Gaussian of $\sigma=0.24$\arcsec
for display, and the F510 image has been smoothed with a $3 \times 3$-pixel (1.2\arcsec) median filter.
North is at the top and east to the left. The cloud is marked by a circle of radius 18 \arcsec in the middle panel. The spectrograph 
slit used for the data in Fig. \ref{fig-eelrspectra} passed through the nucleus of NGC 232, at 
lower right, and the cloud, along position angle $54^\circ$. } 
\label{fig-ngc0235}
\end{figure}

\begin{figure*}
\includegraphics[width=135.mm,angle=90]{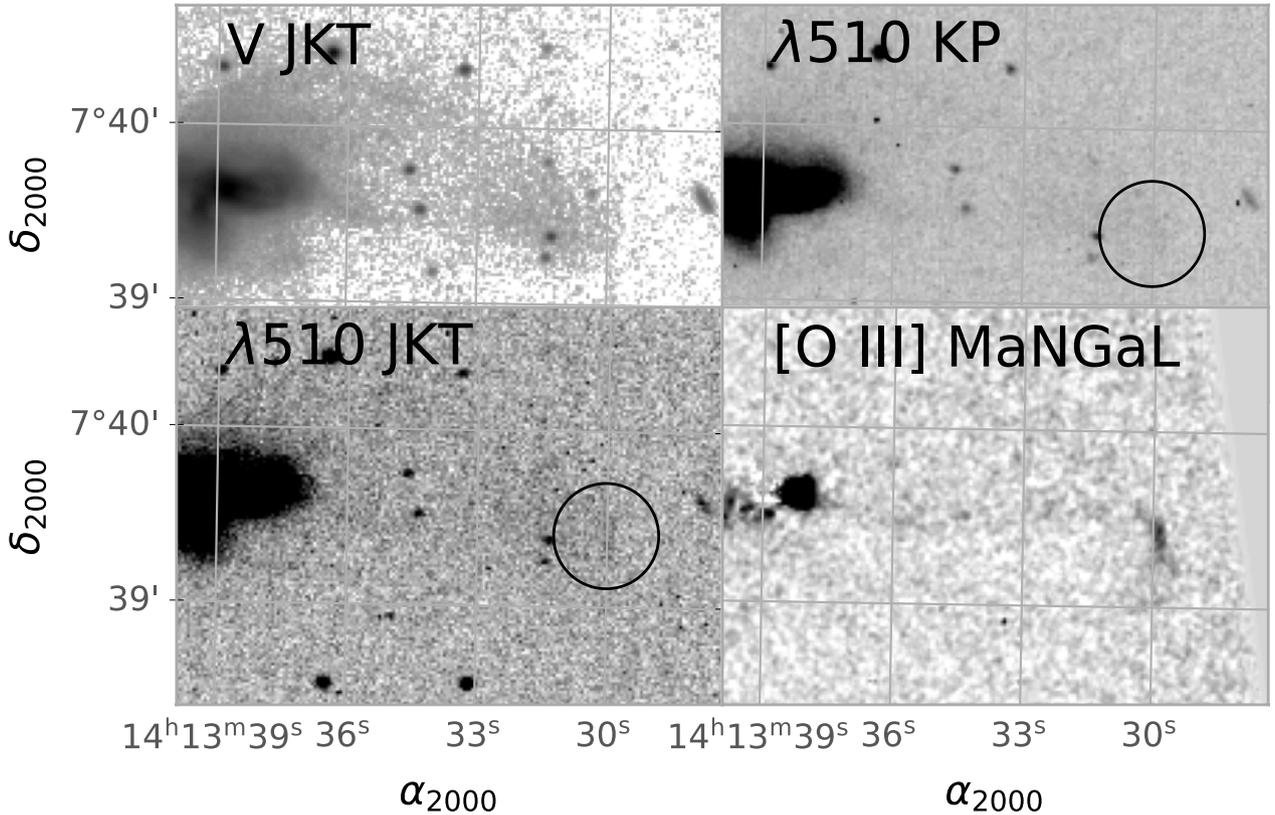}
\caption{The distant emission-line cloud in NGC 5514. The figure compares an offset-log-scaled $V$-band image to show the stellar
structure of this merger with three linearly-scaled views of the cloud - from Kitt Peak (KP), the Jacobus Kapteyn Telescope (JKT) and the
MaNGaL system on the Caucasus 2.5m (MaNGaL). Continuum-subtracted data are denoted as [O III]. The cloud location is 
shown on the two narrowband SARA images, within circles of radius 18 \arcsec. } 
\label{fig-ngc5514}
\end{figure*}

Fig. \ref{fig-ngc0235} shows the cloud in the NGC 235 system, including detection of a portion of the structure in an archival
HST image. The emission-line cloud found in NGC 5514 is especially distant from the nucleus - 135\arcsec (75 kpc) in projection. The cloud
location and structure are shown from three different data sets in Fig. \ref{fig-ngc5514}. In this case, the MaNGaL [O III] data suggest
additional emission closer to the AGN in the same direction as the distant cloud, projected at a radius of 48\arcsec (27 kpc).

Table \ref{tbl-specfits} summarizes the spectroscopic results for the new objects. As in \cite{Mkn1}, errors in
line flux were derived from Gaussian fits to groups of emission lines, including fictitious lines placed in continuum regions, 
constrained to have the same width as [O III] (in the blue) or H$\alpha$ (for red lines). He II is weak, and while formally detected 
at levels 2.1-2.8$\sigma$. These values give a realistic estimate of the S/N ratio of the detections, since the expected 
wavelength is independently known (in both cases the peak appears
within 1 \AA\  of the expected wavelength). Therefore, there is no additional uncertainty such as exists when a peak is sought within
a large range of noisy data (the ``look-elsewhere effect").
As shown in Fig. \ref{fig-bptplots}, these fall into the AGN-ionized region of the primary BPT line-ratio diagrams,
an assessment strengthened by the detections of He II emission at levels characteristic of AGN photoionization.
The distant cloud in NGC 5514 is close to the maximum-starburst  boundary in the first BPT diagram, but the strength of He II
is consistent with AGN rather than hot stars as photoionizing sources. This interpretation fits with the spatial offset 
between the emission-line
cloud and starlight in the adjacent tidal tail, and lack of any obvious star-forming complexes nearby.

Detection of these two distant EELR clouds is the primary result of this study. We consider in section 5 the implications of their
properties of the luminosity history of these AGN, and in section 6 we update the statistics on occurrence of distant EELRs in tidally 
distorted hosts, strengthening the connection between the two phenomena.

\begin{figure*}
\includegraphics[width=180.mm,angle=0]{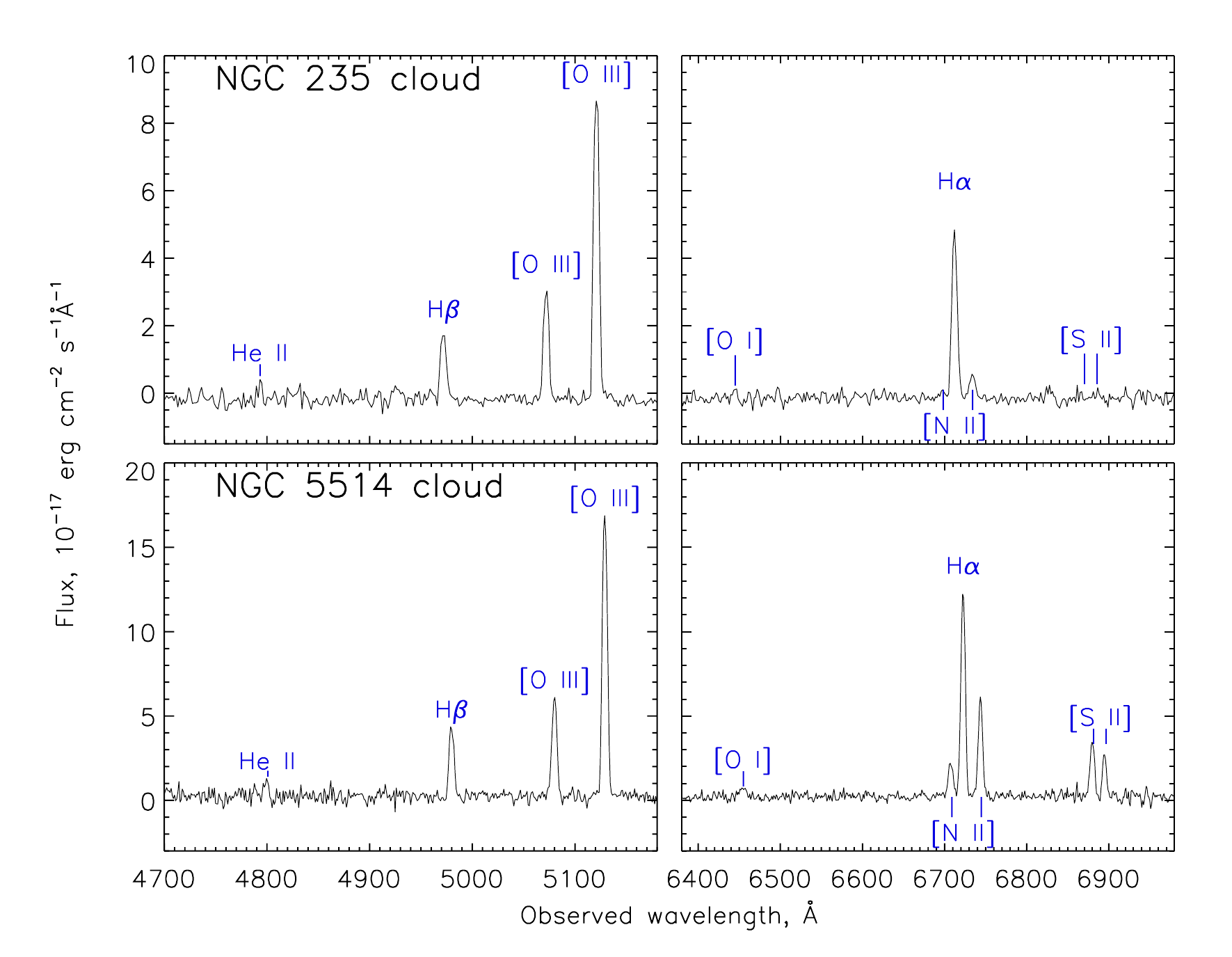}
\caption{BTA-SCORPIO spectra of the two newly identified distant EELRs from this survey.
They are summed over regions $2 \times 8.6$ \arcsec$^2$ for NGC 235 and $1 \times 17.5$\arcsec$^2$ for NGC 5514. Wavelengths 
are in the observed frame; NGC 235 was observed at large airmass, so the telluric B band near 6867 \AA\  is deep and 
could not be well corrected, compromising measurement of the [S II] doublet.} 
\label{fig-eelrspectra}
\end{figure*}

\begin{figure*}
\includegraphics[width=125.mm,angle=90]{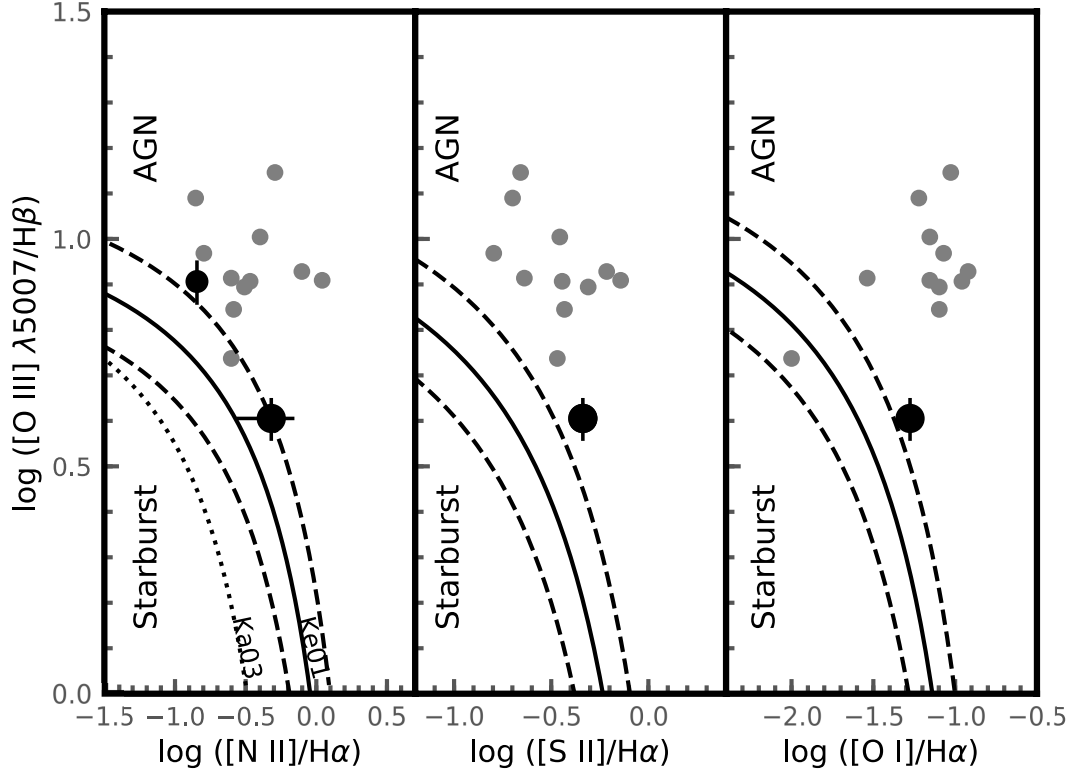}
\caption{Strong-line BPT diagrams, showing the line ratios from the summed spectra of the two new EELRs as in Fig. \ref{fig-eelrspectra}.  
In the first diagram, the boundary between
ionization by hot stars and AGN is shown (as given both by \protect{\citealt {Kewley2001}}, with quoted systematic uncertainty shown by the
dashed curves, and by \protect{\citealt{Kauffmann}}).  Gray symbols indicate the averaged spectra of EELR clouds found more than 10 kpc
from the nuclei in the Galaxy Zoo survey \protect{\citep{Keel2012}}, and the cloud near Mkn 1 \protect{\citep{Mkn1}}. The two newly-found
clouds are indicate by black symbols, larger for the more-distant cloud in NGC 5514, with uncertainty ranges from
Table \ref{tbl-specfits}, some lost within the plotted symbols. Uninformative limits for NGC 235 are not shown. To reduce clutter, 
we do not plot the 
LINER/Seyfert boundary, often taken at log [O III]/H$\beta=0.30$, which passes below all these data points.  [N II] denotes the 6584-\AA\ line, [O I] indicates the 6300-\AA\  line, and [S II] indicates the sum of the 6717 and 6731-\AA\ lines.}
\label{fig-bptplots}
\end{figure*}

\begin{table}
	\centering
	\caption{EELR emission-line properties}
	\label{tbl-specfits}
	\begin{tabular}{lcc}
		\hline
\ Property & NGC 235 & NGC 5514 \\
\ Center RA (2000) & 0:42:49.71 & 14:13:29.88  \\
\ Center Dec (2000) & -23:33:05.1 & +7:39:23.4 \\
\ [O III] $\lambda 5007$ flux in slit (ergs cm$^{-2}$ s$^{-1}$) & $6.75 \times 10^{-16}$ & $9.82 \times 10^{-16}$ \\
\ [O III] $\lambda 5007$/H$\beta$ & $8.06 \pm 0.83$ & $4.03 \pm 0.4$ \\
\ He II $\lambda 4686$/H$\beta$ & $0.21 \pm 0.10$ & $0.22 \pm 0.08$ \\
\ [N II] $\lambda 6583$/H$\alpha$ & $0.143\pm 0.026$ & $0.48 \pm 0.02$ \\
\ [O I] $\lambda 6300$/H$\alpha$ & $<0.042$ & $0.053 \pm 0.006$ \\
\ [S II] $\lambda \lambda 6716, 6731$/H$\alpha$ & --- & $0.46 \pm 0.03$ \\
		\hline
	\end{tabular}
\end{table}

\section{Ionized gas within the AGN hosts - EELRs and ionization cones}

\begin{figure*}
\includegraphics[width=160.mm,angle=0]{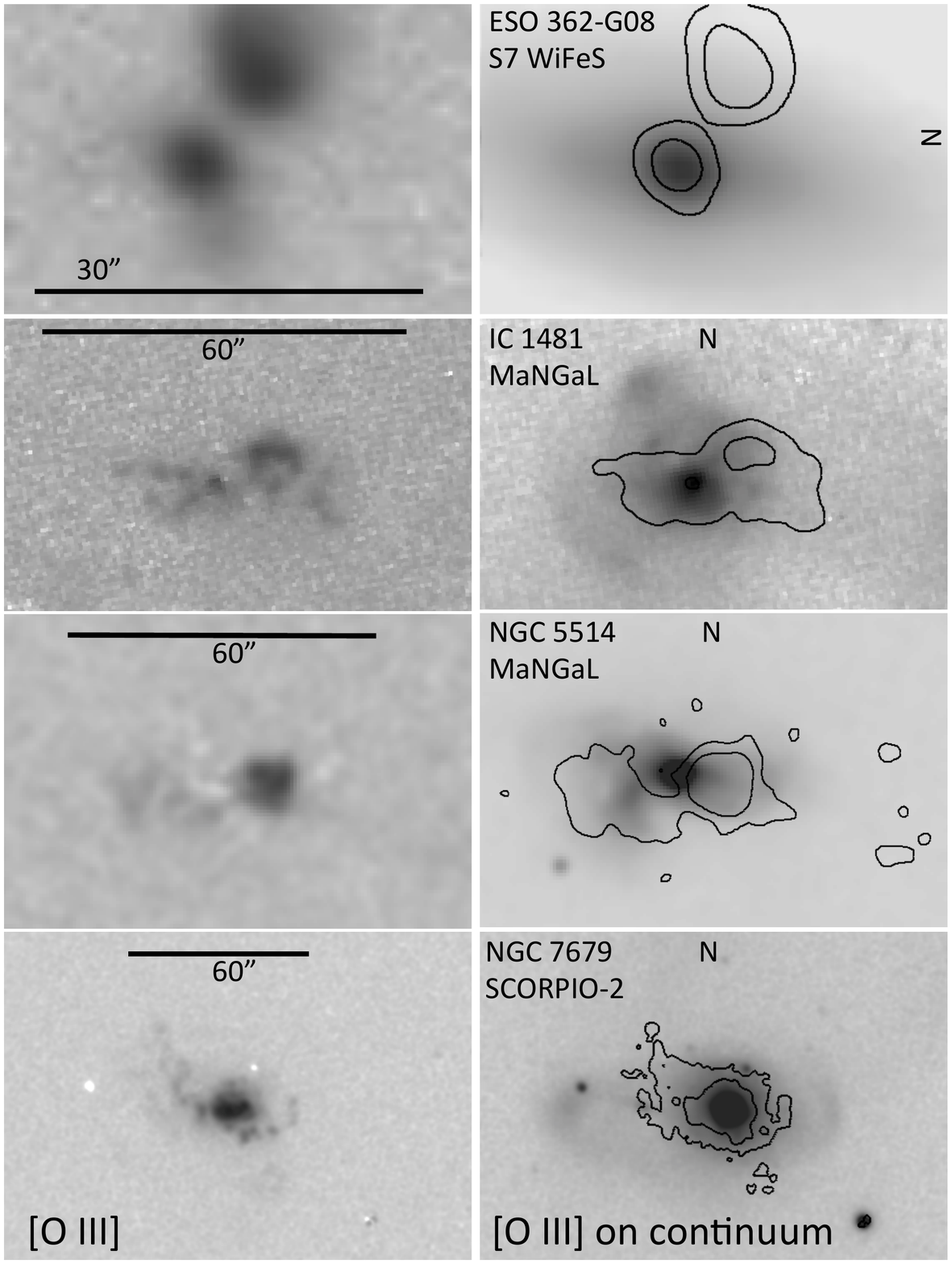}
\caption{Montage of extended emission structures within AGN hosts, from the indicated instruments. The left panels show continuum-subtracted [O III] $\lambda 5007$ emission, while the right panels show a gray-scale mapping of the adjacent continuum, with sparse contours of [O III] superimposed to show their spatial relationship. The intensity scales have been adjusted to show structure over wide dynamic ranges.
Each shows emission over narrow enough angular
spans about the AGN to be candidate ionization cones, even around the low-ionization nuclei of NGC 5514 and IC 1481.
The S7 data for ESO 362-G08 have been sampled to a smoother grid of $3 \times 3$ subpixels for display. North is at the top 
as shown except for ESO 362-G08, where it is to the right.} 
\label{fig-eelrmontage}
\end{figure*}

Our images reveal four systems with extensive emission-line structures having most or all of the detected emission smaller than our 
adopted
cutoff for distant clouds at 10 kpc projected radius, but still AGN-related and spanning much of the host-galaxy ISM. Even with PSF-related artifacts at many nuclei,
these regions were not subtle detections in the survey imaging data, so they are not just the tail of a continuum of EELR sizes and surface brightness. 
We identify likely ionization cones in 
IC 1481 (UGC 12505, CGCG 406-064), recover the bright off-nuclear feature in ESO 362-G08 noted by \citet{Mulchaey}, and 
detect the extensive ionized regions in NGC 7679 and NGC 5514. For completeness, we compare these emission regions, those
whose combination of surface brightness and extent led to their detection in our survey images outside the regions
compromised by PSF mismatch (zones of median radius 6\arcsec). Their [O III] morphologies are compared, along with the
starlight continuum, in Fig. \ref{fig-eelrmontage}. We also consider Mkn 938 (NGC 34), where our survey data are sensitive
enough to detect the ionized emission 18\arcsec NW of the brighter nucleus (7.5 kpc), but not the emission along the tidal
tail NNE, within the uncertainties associated with continuum subtraction. The optical spectrum of the nucleus \citep{Kim}
shows low-ionization LINER emission and a poststarburst stellar population, while integral-field mapping \citep{Mkn938BTA}
shows a powerful outflow and suggest a mix of starburst and AGN contributions, but not AGN ionization farther out.

\subsubsection{ESO 362-G08}
Due to its declination, we obtained no new spectroscopic data for ESO 362-G08, but long-slit data crossing the bright off-nuclear
clouds were analyzed by \cite{Fraquelli}, and integral-field spectroscopic data were released as part of the S7 survey 
\citep{S7}. The extranuclear emission-line structure is dominated by a bright cloud to the NE, with internal velocity structure. The
S7 data also show a dimmer cloud to the WNW; the two clouds might form parts of a set of ionization cones if their
boundaries are defined by the location of otherwise neutral gas. The NE cloud has line ratios putting it firmly in the
AGN region of the usual BPT diagrams, as well as He II/H$\beta = 0.47$.

\subsubsection{NGC 7679} 
\begin{figure*}
\includegraphics[width=170.mm,angle=0]{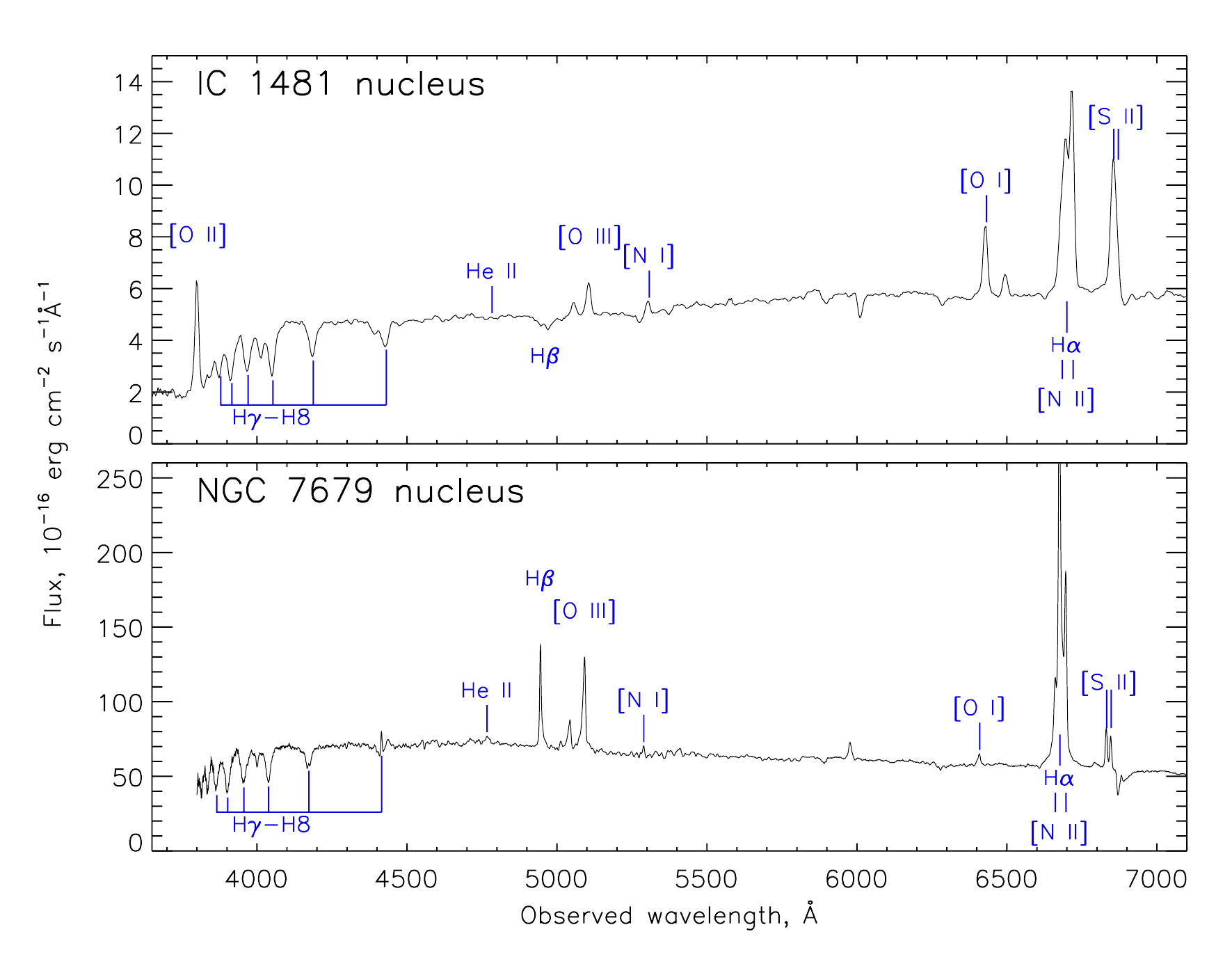}
\caption{BTA SCORPIO-2 spectra of the low-ionization nuclei in IC 1481 and NGC 7679. Each is summed by 2.5\arcsec along the
slit. Important spectral features are marked; the high-order Balmer lines, seen in absorption from a poststarburst population
of stars, are marked by the linked indicators at lower left.} 
\label{fig-IC1481NGC7679spec}
\end{figure*}

While the search results for distant clouds near NGC 7679 were reported in \cite{Mkn1}, we discuss its internal gas emission here
since these were not considered in the earlier paper and we have new relevant data. The nucleus itself may be better described as a
LINER than a Seyfert 2 based on the usual BPT diagrams, in particular having [O III]/H$\beta < 1$ (Fig. \ref{fig-IC1481NGC7679spec}), but He II $\lambda 4686$ is 
relatively strong not only at the nucleus but throughout much of the extended emission region. Like IC 1481, the stellar population
shows post-starburst or quenching signatures, with the higher-order Balmer lines prominent in absorption. The [O III] structure
is elongated, but if it includes ionization cones, there must be a substantial additional contribution near the nucleus broadening its
central area.

Our slit serendipitously crossed a background spiral galaxy, SDSS J232848.57+033042.2, projected 286\arcsec due east of the 
nucleus, at redshift $z=0.051$, whose structure is well resolved in an archival HST image from program 15446. 

\subsubsection{IC 1481}
IC 1481 is listed as showing extraplanar diffuse ionized gas based on the CALIFA IFU survey by \cite{LopezCoba}.
Our BTA SCORPIO spectrum (Fig. \ref{fig-IC1481NGC7679spec}) shows LINER emission at the nucleus, much like NGC 1052, 
with unusually strong [N I] $\lambda 5199$ emission. 
The stellar population shows post-starburst signatures, with strong absorption in the higher-order Balmer lines. 
Higher ionization levels are seen off the nucleus. While the MaNGaL emission-line
image has better spatial sampling and image quality in [O III], the CALIFA data are in good agreement with the overall
emission-line structure, and indicate that both H$\alpha$ and [O III] occur in similar regions (which may mean that the structure
we see is limited by the presence of neutral gas rather than ionizing photons). The CALIFA radial-velocity measures
show little systematic motion across the emission structures. 

The morphology of the emission-line structure can be interpreted as a pair of ionization cones
with axes east-west, crossing very close to the continuum and H$\alpha$ peak, so IC 1481 may join NGC 1052 \citep{NGC1052} and NGC 2442 \citep{NGC2442} as LINERs with ionization cones.
These objects implicitly show that AGN with low radiative output can still be surrounding by an obscuring torus, and may
be fruitful in investigations of how the AGN output and surrounding gas structure interact to produce low-ionization
line emission coupled with significant He II emission, constrained by the X-ray spectra.

The MaNGAL data were obtained after our long-sit spectrum, so the spectrum was aligned with the brightest off-nuclear knot to the northwest (including the spectrum of scattered light from the 8th-magnitude K star HD 219905 to the SE) . As shown in
Fig. \ref{fig-eelrmontage}, IC 1481 shows merger or post-merger signatures; even the faint features extending beyond the
MaNGaL continuum field are detected in both SARA and Legacy Survey images as well.

\subsubsection{NGC 5514}
\begin{figure*}
\includegraphics[width=135.mm,angle=0]{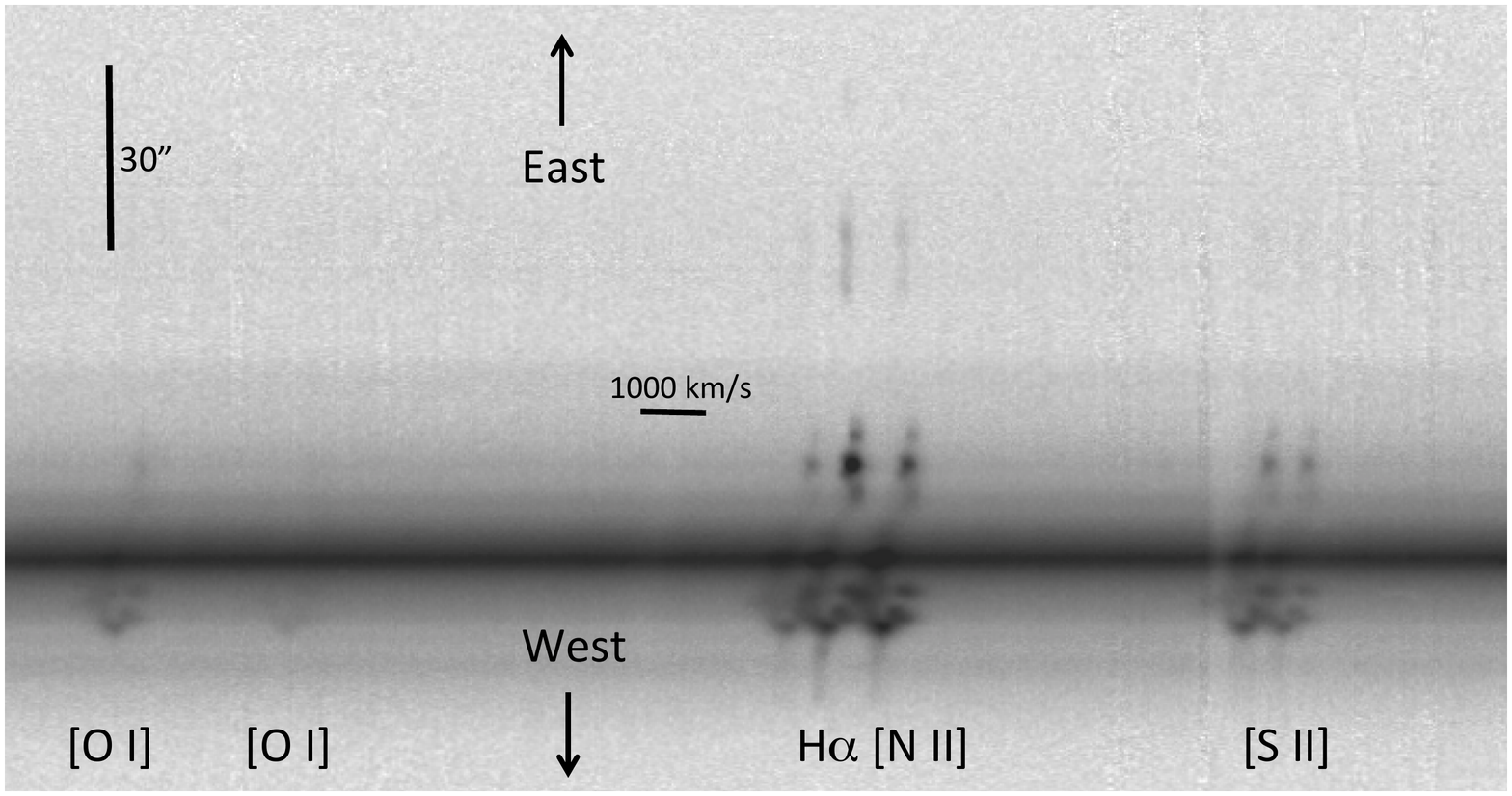}
\caption{Long-slit spectrum of the central regions of NGC 5514, along position angle (PA) 88$^\circ$. This section at
red wavelengths shows the patchy elliptical traces of an expanding shell west of the nucleus. The velocity scale applies strictly 
at H$\alpha$.  } 
\label{fig-ngc5514grayscale}
\end{figure*}

\begin{figure*}
\includegraphics[width=135.mm,angle=0]{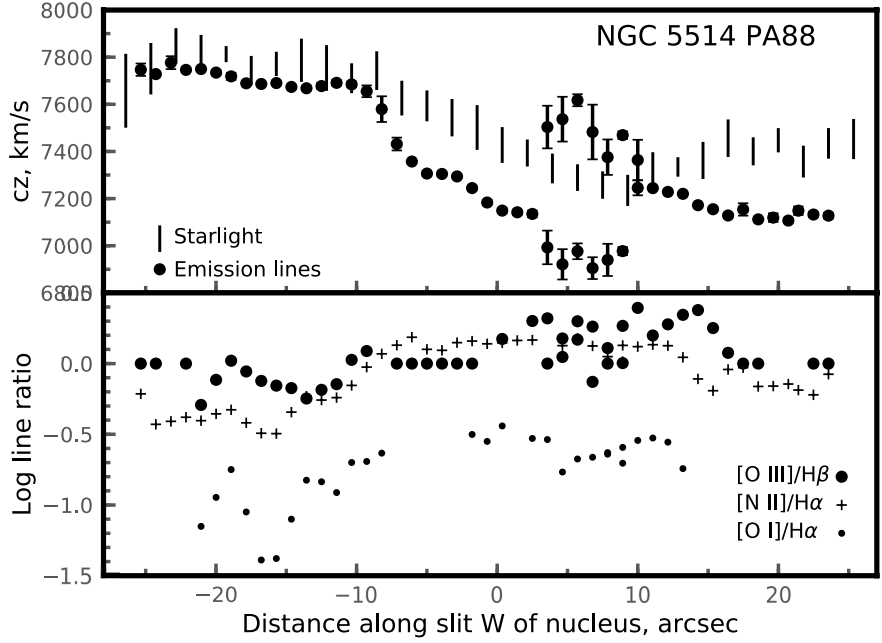}
\caption{Spectroscopic properties of NGC 5514 along our slit in position angle 88$^\circ$. Top, emission-line velocities, averaging over
lines detected and unconfused by B-band absorption or other velocity components at each position, with the internal standard
deviation shown at each position (filled circles and error bars), compared to the cross-correlation stellar velocities (shown as
$\pm 1 \sigma$ error ranges). 
For measurement, data were summed along 3-pixel (1.05\arcsec ) regions along the slit. Lower panel, important emission-line
ratios along the slit, plotted as the decimal logarithm as used in the BPT diagrams. Virtually all these locations fall within the
``LINER" regions of the diagrams, interpreted by \citet{Lipari} as showing a dominant role for shock ionization in most of the
main body of NGC 5514. We observe no increased ionization near the optically brighter northwestern nucleus, which is 
at the origin of the angular scale shown.} 
\label{fig-ngc5514slit}
\end{figure*}

\begin{figure}
\includegraphics[width=64.mm,angle=90]{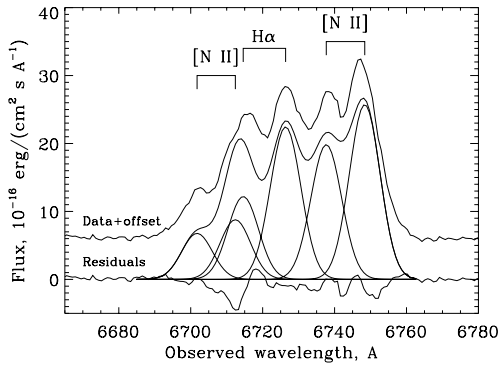}
\caption{Sample Gaussian decomposition of the blended H$\alpha$-[N II] emission lines for a 1.05\arcsec region near the greatest velocity 
separation of components, as used in Fig. \ref{fig-ngc5514slit}. The components were constrained to have the same width, and each velocity component of 
the weaker [N II]$\lambda 6548$  line
was constrained to have ratio 1:3 in photon rate to the stronger $\lambda 6583$ line. The bluer H$\alpha$ and redder [N II] components
nearly overlap at this location, as do components of the redder [S II] doublet (making the [S II] line ratio in the components almost unconstrained). The data are offset upward for clarity; curves show individual model components and their sum. The residuals from
this fit are shown at the bottom, illustrating the limitations of a Gaussian line profile in matching the line shapes in detail.} 
\label{fig-ngc5514fit}
\end{figure}

\begin{figure*}
\includegraphics[width=140.mm,angle=90]{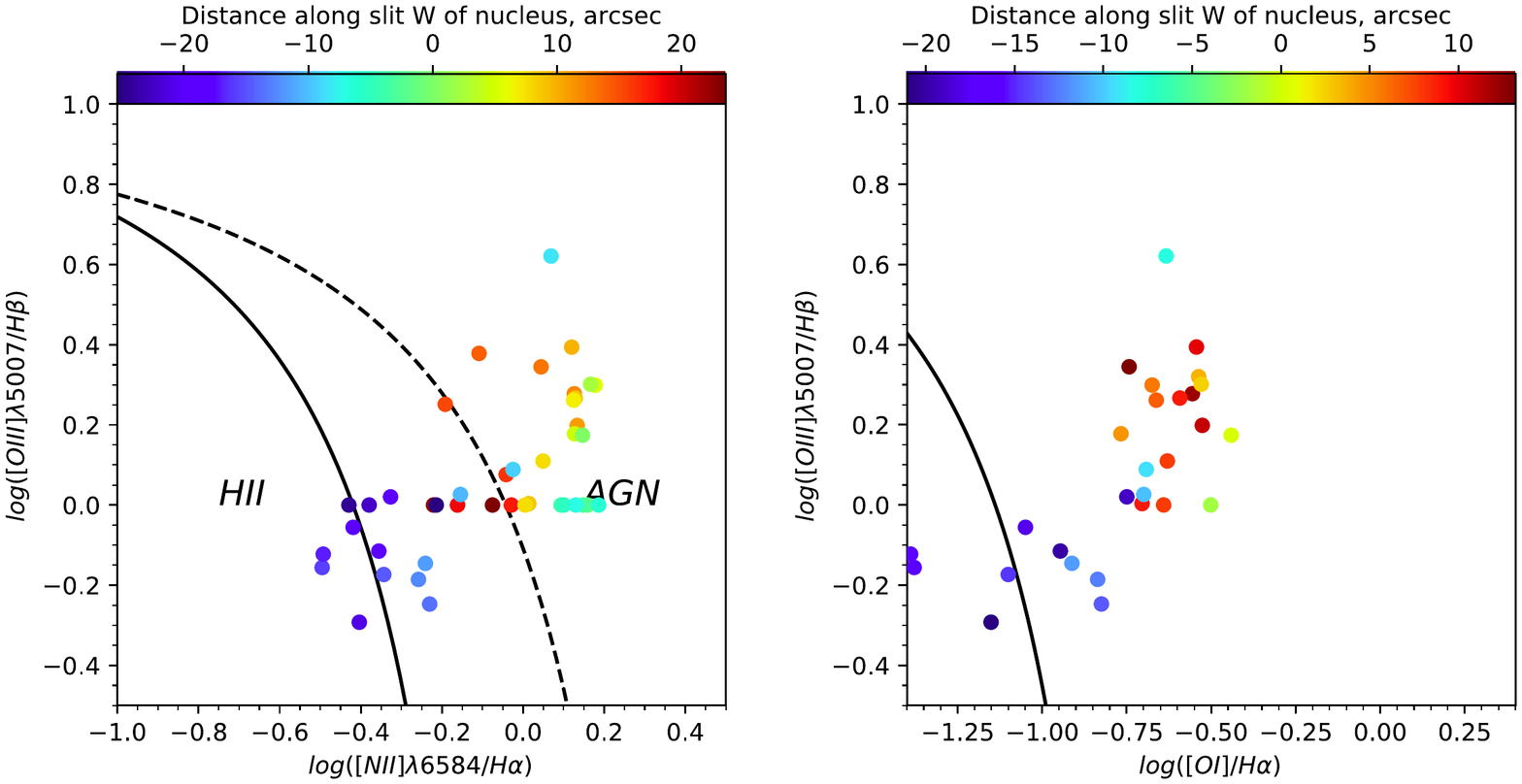}
\caption{BPT diagrams of [O III]/H$\beta$ versus [N II]/H$\alpha$ and [O I]/H$\alpha$for the long-slit data in Fig. \ref{fig-ngc5514slit}, 
color-coded with location along the slit according to the color bar at the top. The relative contribution of shocks (or AGN emission) increases broadly going to the west side of the system, with large local scatter. Regions denoting emission from star-photoionized H II regions and AGN (or shocks) are indicated.As before, [N II] denotes the 6583-\AA\ line and [O I] indicates the 6300-\AA\  line. } 
\label{fig-NGC5514BPT1}
\end{figure*}

\cite{Lipari} have used integral-field spectroscopy to analyze the kinematics and ionization of the gas in the central regions of
NGC 5514 in great detail, finding multiple expanding (super)bubbles, and attributing the bulk of the ionization to the accompanying shocks.
Our long-slit spectrum (Fig. \ref{fig-ngc5514grayscale}) shows complex emission structure, with some very distant components where the slit crosses
tidal tails, and double line profiles in a pattern suggesting an expanding shell on the western side. We examine this
structure through multicomponent Gaussian fits to emission lines and line blends, done with the {\it splot} task in IRAF,
in summed 3-pixel (1.05\arcsec) steps, after reducing the effects of absorption features from starlight via
subtracting a template spectrum at the radial velocity derived from cross-correlation at each point along the slit. These
results are illustrated for a region with the largest velocity separation in Fig. \ref{fig-ngc5514fit}.
The line profiles by themselves are double-peaked when the separation is
large compared to the line width, so what happens at locations where Fig. \ref{fig-ngc5514grayscale} shows the lines to be just barely splitting is
poorly constrained. In assessing the gas kinematics, we omit lines where components of adjacent lines (such as H$\alpha$ and one 
of the adjacent [N II] features) nearly overlap at a particular slit
location, since the measured wavelength will reflect blending. H$\beta$, [O III] $\lambda 5007$, and [O I] $\lambda 6300$
are free of this ambiguity. Fig. \ref{fig-ngc5514slit} compares mean values of the emission-line velocities of each component with cross-correlation
stellar velocities along the slit, providing a complementary view to the maps shown by \cite{Lipari}. The blue- and redshifted components
are separated by as much as 650 kms$^{-1}$, and the double line profiles are roughly symmetric about the stellar velocities. The redder
component departs from the stellar velocity systematically at points to the west of about 9\arcsec east of the nucleus, suggesting that 
this large expanding structure is seen substantially projected across the core. The slopes of the emission-line velocities compared to the
stellar kinematics further suggest that this component is expanding away from us. The emission-line ratios, and implied mix of
ionization sources, change in ways that correlate only poorly with location (Fig. \ref{fig-NGC5514BPT1}).

Table l\ref{tbl-the4eelrs} compares the properties of these four extended emission regions, measured from our data and the archival
results from the S7 survey \citep{S7}. We attempt to separate the [O III] component from the ``normal" narrow-line region (NLR) spatially,
taking the NLR to be the central peak similar in scale to the PSF. For these galaxies, the total [O III] luminosity is dominated by the
EELR, providing additional examples of the situation noted by \cite{Hainline2014} and \cite{Hainline} where large regions of the host galaxies need to
be included in measurements of emission lines, such as [O III], to derive an accurate total luminosity for the AGN. This holds 
even for the relatively low AGN luminosities we deal with here. (These luminosities include foreground Galactic extinction values 
following \citealt{Schlafly}). For ESO 362-G04, some of the cloud to the NE may fall outside the S7 integral-field data and
be excluded from the listed [O III] flux. In contrast, much of the emission in both regions in NGC 5514 can be attributed to
shocks associated with outflows, and the central emission is weak and simply summed within 3\arcsec of the continuum peak. 
Both [O III] luminosities in NGC 5514 are therefore upper limits to the AGN contribution.

\begin{table*}
	\centering
	\caption{EELR emission-line properties.}
	\label{tbl-the4eelrs}
	\begin{tabular}{lcccc} 
		\hline
\ Object	& ESO 362-G04 & IC 1481	 & NGC 5514 & NGC 7679\\
\ AGN type &      Sy 2          & LINER          & LINER        & Sy 2 \\
\ $z$          &     0.016          &  0.020          &   0.024         & 0.017 \\         
\ [O III] $\lambda 5007$ NLR flux (ergs cm$^{-2}$ s$^{-1}$) & $3.0 \times 10^{-15}$ & $2.6 \times 10^{-15}$ & $1.1 \times 10^{-15}$ & $6.6 \times 10^{-14}$ \\
\ [O III] $\lambda 5007$ EELR flux (ergs cm$^{-2}$ s$^{-1}$) & $ 2.3 \times 10^{-15}$ & $2.0 \times 10^{-14}$ & $1.8 \times 10^{-13}$ & $4.4 \times 10^{-13}$ \\
\ [O III] $\lambda 5007$ NLR luminosity (ergs s$^{-1}$) & $2.2 \times 10^{39}$ & $2.9 \times 10^{39}$ & $1.8 \times 10^{39}$ & $5.1 \times 10^{40}$ \\
\ [O III] $\lambda 5007$ EELR luminosity (ergs s$^{-1}$) & $1.7 \times 10^{39}$ & $3.7 \times 10^{40}$ & $3.0 \times 10^{41}$ & $3.4 \times 10^{41}$  \\
\ NLR/total [O III] & 0.56 & 0.07 & 0.006 & 0.11\\
\ EELR radii (\arcsec) & 3--17 & 3--23 & 3--22 & 1.5--11 \\
\ Cone opening angle ($^\circ$) & 109 & 70 & 51 & 60 \\
		\hline
	\end{tabular}
\end{table*}

\section{Energy balance and AGN luminosity history}

The relevant quantities to examine the energy balance between the AGN as observed directly, and photoionization
requirements for the distant clouds, are shown in Table \ref{tbl-energy}. The recombination-balance luminosity
lower limit follows \cite{Keel2012}, with the value for NGC 5514 adjusted for the fact that the expression they
used assumed a slit oriented radially to the AGN while our slit position was almost perpendicular to that in order to
include more of the cloud. To account for this, estimating what we measure by integrating along a radial slit, 
we multiply the mean H$\beta$ surface brightness within the cloud per arcsecond$^2$ by the radial extent 4.4\arcsec from our [O III] images. We take the FIR luminosity as an upper bound on the absorbed AGN flux, since it is integrated over
essentially the whole galaxy at these distances and may include a substantial contribution from ongoing star formation.
We also consider the mid-IR contribution to the obscured-AGN luminosity, derived from WISE PSF magnitudes as in 
\cite{crossion}. For the unobscured AGN ionizing luminosity, we scale from the [O III] luminosity rather than broad lines 
since these are type 2 AGN, following \cite{lamassa} for a mean relation of observed [O III] to mid-IR luminosity of 1/400, and
mean ratio of mid/IR to ionizing luminosity of 0.8 (following \citealt{QED}), so the unobscured ionizing luminosity is 320 times the 
nuclear [O III] luminosity. Comparison of the total AGN luminosity, an upper bound since not all the FIR emission is
from grains heated by ionizing radiation (with a substantial contribution by young stars possible), to the minimum requirements to power the
extended clouds, shows that NGC 235 is consistent with a constant luminosity over the extra light-travel time from nucleus to 
cloud and then to the observer, while NGC 5514 has a shortfall of at least 2.5 times. Especially for NGC 5514, the shortfall may 
be significant despite the approximate nature of the luminosity analysis, since analysis of EELR images at higher resolution 
\citep{Keel2017b} shows that 
the highest-surface-brightness EELR clouds typically absorb only a fraction $\approx 1/3$ of the incident ionizing radiation.
Furthermore, if much of the ionization near the core of NGC 5514 is by shocks rather than photons from the AGN \citep{Lipari}, the
implied AGN luminosity as observed becomes an upper limit.

NGC 235 is part of a multiple system including NGC 232, a comparably luminous spiral in which \cite{NGC232} have detected
a radial, high-ionization emission line-structure extending 3 kpc in projection along position angle $163^\circ$, likely related to
an AGN on this galaxy. The EELR we have found is seen roughly midway between NGC 235 and NGC 232; the nucleus of
NGC 232 has lower ionization than NGC 235, very close to the LINER/Seyfert boundary from \cite{NGC232}, so its estimated
ionizing output is correspondingly lower. We therefore attribute the ionization on the cloud to the brighter AGN in NGC 235.

As noted above and by \citet{Lipari}, the optical emission from gas in NGC 5514 has low ionization, and not line ratios and the
complex kinematics are consistent with widespread shock ionization. in fact, it is unclear whether our spectral data allow us to isolate an AGN component in this system, motivating us to
look more broadly for estimates of AGN luminosity as observed directly. Conservatively, the lack of changes in ionization-sensitive
line ratios such as [O III]/H$\beta$ and [O I]/H$\alpha$  between the nucleus and its surroundings (Fig. \ref{fig-ngc5514slit})
suggests that less than half the
emission projected at the nucleus could come from a higher-ionization typical AGN component. No high-sensitivity pointed X-ray observations have been obtained; the typical flux level $2 \times 10^{-13}$ ergs cm$^{-2}$ s$^{-1}$ for the ROSAT survey
\citep{Boller} would translate to ionizing luminosity $\approx 3 \times 10^{41}$ ergs s$^{-1}$ at this distance. Any spectral signature of 
the AGN must fall with the  ``obscured" category, based on far-IR data. An important role for dust heating by star formation is likely in
dusty merging systems of this kind.

\begin{table*}
	\centering
	\caption{Quantities related to the energy balance between AGN output, as seen directly, and photoionization requirements to power the distant ionized clouds.}
	\label{tbl-energy}
	\begin{tabular}{lcc} 
		\hline
Object & NGC 235 & NGC 5514 \\
$z$ & 0.0222 & 0.0235\\
Nucleus [O III] flux, ergs cm$^{-2}$ s$^{-1}$ & $1.18 \times 10^{-13}$ & $9.4 \times 10^{-16}$  \\
Nucleus [O III] luminosity, ergs s$^{-1}$ & $1.3 \times 10^{41}$ & $1.1 \times 10^{39}$  \\
Unobscured ionizing luminosity, ergs s$^{-1}$ & $4.2 \times 10^{43}$ & $3.5 \times 10^{41}$ \\ 
MIR flux,  erg cm$^{-2}$ s$^{-1}$ & $1.0 \times 10^{-10}$ & $6.0 \times 10^{-12}$  \\
MIR luminosity, erg s$^{-1}$  & $2.4 \times 10^{40}$ & $1.2 \times 10^{39}$ \\
FIR flux,  erg cm$^{-2}$ s$^{-1}$  &  $1.4 \times 10^{-6}$ & $7.2 \times 10^{-7}$ \\
FIR luminosity erg s$^{-1}$ & $2.4 \times 10^{44}$ & $1.4 \times 10^{44}$ \\
Obscured+unobscured ionizing luminosity,  ergs s$^{-1}$ & $2.8 \times 10^{44}$ & $1.4 \times 10^{44}$ \\
Cloud H$\beta$ flux, erg cm$^{-2}$ s$^{-1}$ & $1.5 \times 10^{-16}$ & $5.4 \times 10^{-17}$ \\
Projected separation, arcsec & 57 & 135 \\
Projected separation, kpc & 26 & 75  \\
Recombination-balance AGN luminosity, erg s$^{-1}$  & $>2.4  \times 10^{44}$ & $>3.9 \times 10^{44}$  \\
		\hline
	\end{tabular}
\end{table*}

\section{Distant EELRs and tidal debris}

\begin{figure*}
\includegraphics[width=135.mm,angle=0]{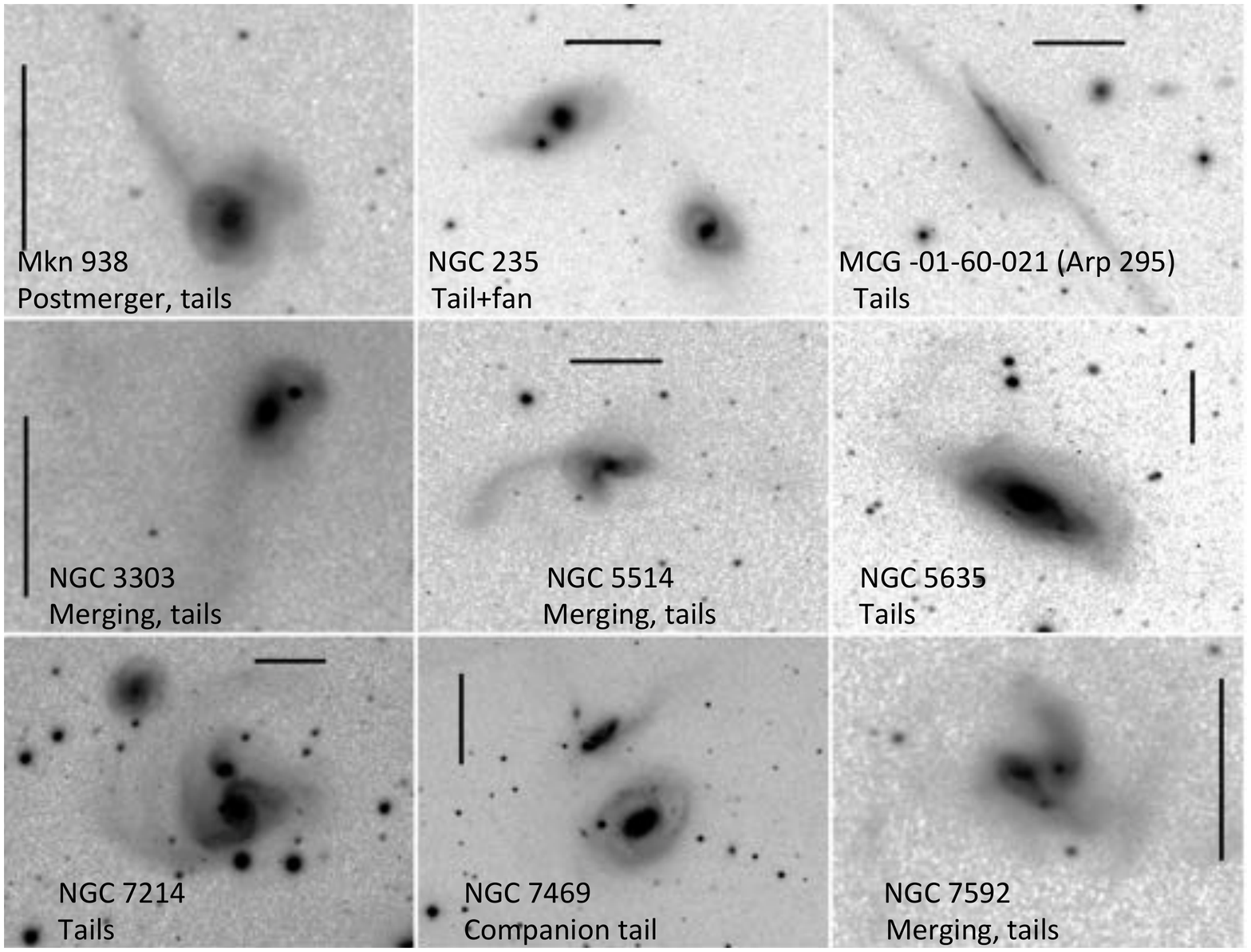}
\caption{Sample images of kinds of tidal features in Table \ref{tbl-tails}, from SARA $V$-band images. Scale bars show 1 arc minute in each case. The intensity mapping was tailored for each image, to show both bright structures and the faint tidal features.} 
\label{fig-tails}
\end{figure*}

The kinematically quiescent distant EELRs studied in our earlier work (\citealt{Voorwerp}, \citealt{Keel2012}, \citealt{Mkn1})
occur overwhelmingly in systems which are strongly interacting or merger remnants, as are the two new distant EELRs 
identified in this survey. As a step toward using these results to estimate the occurrence of these large-scale
ionized clouds around AGN, we have examined our broadband images and the Legacy Survey data \citep{LegacySurveys} in
the VCV sample for evidence of 
tidal effects - near-radial tails, tangential shells, and external loops or arcs (Table \ref{tbl-tails}). ``Data source" indicates whether the listed
tidal structure was detected on our SARA $V$ images or seen only on Legacy Survey $g$ images (when both sources are
listed, they apply to different structures in order). The surface-brightness thresholds are bimodal, since not all
sample galaxies are included in the Legacy Survey (and the Legacy Survey differs in exposure properties essentially between northern and southern regions). Typical detection thresholds of the structures noted here are
$V=25.6$ \arcsec$^{-2}$ for SARA, very roughly three times deeper or $g\approx 27$ \arcsec$^{-2}$ for the Legacy Survey.
We see tidal structures - tails, shells, or arcs  - in 26 of the 111 VCV sample galaxies (23\%); our categories are illustrated
with SARA $V$ images in Fig. \ref{fig-tails}. Very broadly, we detect distant 
EELR emission in 10\% of disturbed AGN hosts. One occurs in the Toomre sample (albeit marginally reaching our
10-kpc radial limits), which was preselected for tidal disturbances. 

All the EELRs detected in this survey are sections of tidal debris. This extends the results from \cite{Keel2012}, where
at least 18 of 19 EELRs were in clearly interacting or merging galaxies (Mkn 78 was the possible exception).

\begin{table}
	\centering
	\caption{Tidal features in sample galaxies from SARA and Legacy Survey (LS) data.}
	\label{tbl-tails}
	\begin{tabular}{lcc} 
		\hline
Galaxy & Structure & Data source\\
ESO 548-G81   & Tails & LS\\
ESO 575-IG016 &  Distorted; tail &  SARA \\
IC 1481         &  Postmerger, tails    & SARA/LS \\
IC 1833        & Patchy shells  & LS \\
Mkn 938     &    Postmerger, tail to 84\arcsec  & SARA \\
NGC 235           & tail+fan   & SARA \\
MCG -01.05.047    & End of disc warped; NNW compn has tail & LS \\
MCG -01.34.008   & External loops  & LS \\
MCG -01.60.021  & Tails  & SARA \\
MCG -03.05.007   & Outer asymm arm or tail     & LS \\
NGC  788         & Shells & LS  \\
NGC 3303A   & Merging, tails & SARA \\
NGC 5318    & Faint tails  & LS \\
NGC 5395     & Tail   & LS \\
NGC 5514      & Merging/tails & SARA \\
NGC 5635          & Faint tails & SARA \\
NGC 6211        & Faint tail  & LS \\
NGC 6240         & merging, tails & SARA \\
NGC 7214     &   Tails & SARA \\
NGC 7319    &  Tail & SARA \\
NGC 7469  &  Compn w/tail;  tail/arc  & SARA/LS \\
NGC 7592E    & Merger, tail & SARA \\
NGC 7592W   &  Merger, tail &  SARA \\
UGC  7064      & Tail & LS  \\
NGC 7679     & merger, tails  & SARA/LS \\
NGC 7682     & Arcs & LS   \\
		\hline
	\end{tabular}
\end{table}

We use starlight features as tracers of tidal distortion strong enough to bring substantial neutral gas outside the
main ISM of a galaxy, and often well outside its inner-disc plane with attendant absorption as well. While H I tails
are typically roughly coincident with stellar tails, they may extend much farther and have some angular offset \citep{HIRogues}.
Unlike the sample in \cite{Mkn1}, H I mapping has not yet been done for most of the galaxies in this sample.

\section{Conclusions}

From an [O III] imaging survey of 17 merging systems and 109 additional putative AGN hosts, we find
three AGN-photoionized distant emission-line regions, projected more than 10 kpc from the galaxy nucleus.
The cloud in the merger remnant NGC 7252 was previously known \citep{NGC7252} and only partly passes our 10-kpc threshold. 
We report new detections, with spectroscopic confirmation, of these distant EELR clouds in the strongly interacting systems 
NGC 235 and NGC 5514. Emission-line ratios indicate that they are photoionized by the AGN. Together with earlier surveys, 
this strengthens that pattern of kinetically quiescent distant EELRs occurring almost solely in interacting or merging systems. Of 
systems with tidal distortion in available images, roughly 10\% show these distant ionized clouds. Energy balance indicates that 
the AGN in NGC 5514 has likely faded by at least a factor 3 over the light-travel time difference of order 250,000 years; the
AGN in NGC 235 may have faded but the uncertainties are too large to make a strong statement.

Our survey images also show four systems with unusually large and luminous EELRs within the host galaxies. These 
range from 1--170 times as bright as the usual NLR in [O III]. We find likely ionization cones in IC 1481, unusual
for a LINER nucleus, and elongated emission on comparable scales around the similarly low-ionization nucleus of NGC 7679.
Both these galaxies show post-starburst stellar populations.

This survey contributes to our understanding of the occurrence of distant AGN-ionized clouds, and of the timescales
of luminous AGN episodes. Use of a narrowband imaging survey reaches lower surface brightness levels than 
colour selection from SDSS images (as in the Galaxy Zoo survey of \citealt{Keel2012}), sampling clouds which
may be farther from the ionizing source and have higher space density. A similar survey specifically of merging systems 
is ongoing, as is colour selection using Legacy Survey data carried out by Galaxy Zoo participants. Together, these
samples will further improve constraints on the radiation escaping AGN in direction and time.

\section*{Acknowledgements}

The narrowband filters were obtained thanks to a University of Alabama College of Arts and Sciences Dean's Leadership Board Faculty Fellowship.
Acquisition of new imagers for the SARA Observatory was supported by the National Science Foundation through
grant 0922981 to East Tennessee State University, and retrofitting of the Jacobus Kapteyn Telescope through grant 
1337566 to Texas A\&M University -- Commerce. We thank the BASS team for public release of their optical spectral data
and fits. We thank Dr. Sebastian Sanchez for access to data obtained as part of the eCALIFA extension of their IFU survey, and
Dr. Adam Thomas for facilitating access to the S7 data release.

This work is partly based on observations obtained with the 6-m telescope of the Special Astrophysical Observatory of the
Russian Academy of Sciences carried out with the financial support of the Ministry of Science and Higher Education of the
Russian Federation (including agreement No. 05.619.21.0016, project ID RFMEFI61919X0016). The work of the 2.5-m telescope is
supported by the Program of development of M.V. Lomonosov Moscow State University.
The analysis of the ionized gas properties according to ADAM, SCORPIO, SCORPIO-2 and MaNGaL data performed by 
the SAO RAS team was
supported by the grant of Russian Science Foundation project 17-12-01335 ``Ionized gas in galaxy discs and beyond the optical radius".
The work of the 1.6-m telescope was performed with budgetary funding of Basic Research program II.16; the telescope is equipment of Center for Common Use ``Angara" (http://ckp-rf.ru/ckp/3056/).

The authors, and the Legacy Surveys project, are honored to be permitted to conduct astronomical research on Iolkam Du'ag (Kitt Peak), a mountain with particular significance to the Tohono O'odham Nation.
This research has made use of the NASA/IPAC Extragalactic Database (NED), which is operated by the Jet Propulsion Laboratory, 
Caltech, under contract with the National Aeronautics and Space Administration.
This research has made use of the VizieR catalogue access tool, CDS, Strasbourg, France (DOI : 10.26093/cds/vizier). The 
original description of the VizieR service was published in 2000, A\&AS 143, 23.

The Legacy Surveys consist of three individual and complementary projects: the Dark Energy Camera Legacy Survey (DECaLS; Proposal ID \#2014B-0404; PIs: David Schlegel and Arjun Dey), the Beijing-Arizona Sky Survey (BASS; NOAO Prop. ID \#2015A-0801; PIs: Zhou Xu and Xiaohui Fan), and the Mayall $z$-band Legacy Survey (MzLS; Prop. ID \#2016A-0453; PI: Arjun Dey). DECaLS, BASS and MzLS together include data obtained, respectively, at the Blanco telescope, Cerro Tololo Inter-American Observatory, NSF's NOIRLab; the Bok telescope, Steward Observatory, University of Arizona; and the Mayall telescope, Kitt Peak National Observatory, NOIRLab. 
NOIRLab is operated by the Association of Universities for Research in Astronomy (AURA) under a cooperative agreement with the National Science Foundation.

This project used data obtained with the Dark Energy Camera (DECam), which was constructed by the Dark Energy Survey (DES) collaboration. Funding for the DES Projects has been provided by the U.S. Department of Energy, the U.S. National Science Foundation, the Ministry of Science and Education of Spain, the Science and Technology Facilities Council of the United Kingdom, the Higher Education Funding Council for England, the National Center for Supercomputing Applications at the University of Illinois at Urbana-Champaign, the Kavli Institute of Cosmological Physics at the University of Chicago, Center for Cosmology and Astro-Particle Physics at the Ohio State University, the Mitchell Institute for Fundamental Physics and Astronomy at Texas A\& M University, Financiadora de Estudos e Projetos, Fundacao Carlos Chagas Filho de Amparo, Financiadora de Estudos e Projetos, Fundacao Carlos Chagas Filho de Amparo a Pesquisa do Estado do Rio de Janeiro, Conselho Nacional de Desenvolvimento Cientifico e Tecnologico and the Ministerio da Ciencia, Tecnologia e Inovacao, the Deutsche Forschungsgemeinschaft and the Collaborating Institutions in the Dark Energy Survey. The Collaborating Institutions are Argonne National Laboratory, the University of California at Santa Cruz, the University of Cambridge, Centro de Investigaciones Energeticas, Medioambientales y Tecnologicas-Madrid, the University of Chicago, University College London, the DES-Brazil Consortium, the University of Edinburgh, the Eidgenossische Technische Hochschule (ETH) Zurich, Fermi National Accelerator Laboratory, the University of Illinois at Urbana-Champaign, the Institut de Ciencies de l'Espai (IEEC/CSIC), the Institut de Fisica de Altes Energies, Lawrence Berkeley National Laboratory, the Ludwig Maximilians Universitat Munchen and the associated Excellence Cluster Universe, the University of Michigan, NSF's NOIRLab, the University of Nottingham, the Ohio State University, the University of Pennsylvania, the University of Portsmouth, SLAC National Accelerator Laboratory, Stanford University, the University of Sussex, and Texas A\& M University.

BASS is a key project of the Telescope Access Program (TAP), which has been funded by the National Astronomical Observatories of China, the Chinese Academy of Sciences (the Strategic Priority Research Program ``The Emergence of Cosmological Structures" Grant \#XDB09000000), and the Special Fund for Astronomy from the Ministry of Finance. The BASS is also supported by the External Cooperation Program of Chinese Academy of Sciences (Grant \# 114A11KYSB20160057), and Chinese National Natural Science Foundation (Grant \# 11433005).

The Legacy Survey team makes use of data products from the Near-Earth Object Wide-field Infrared Survey Explorer (NEOWISE), which is a project of the Jet Propulsion Laboratory/California Institute of Technology. NEOWISE is funded by the National Aeronautics and Space Administration.

The Legacy Surveys imaging of the DESI footprint is supported by the Director, Office of Science, Office of High Energy Physics of the U.S. Department of Energy under Contract No. DE-AC02-05CH1123, by the National Energy Research Scientific Computing Center, a DOE Office of Science User Facility under the same contract; and by the U.S. National Science Foundation, Division of Astronomical Sciences under Contract No. AST-0950945 to NOAO.

\section*{Data Availability}

 The new data described here have been deposited with zenodo.org under DOI 10.5281/zenodo.5248363.











\bsp	
\label{lastpage}
\end{document}